\documentstyle[aps,eqsecnum,epsfig]{revtex} 
\begin{document} 
\title{\bf Damping Rates and Mean Free Paths of Soft Fermion Collective Excitations 
in a Hot Fermion-Gauge-Scalar Theory\footnote{To appear in Physical Review D}}   
\author{\bf  S.-Y. Wang$^{(a)}$, D. Boyanovsky$^{(a,b)}$, H.J. de Vega$^{(a,b)}$, 
D.-S. Lee$^{(c)}$, and Y.J. Ng$^{(d)}$}
\address
{(a) Department of Physics and Astronomy, University of Pittsburgh, 
Pittsburgh, Pennsylvania 15260\\ 
(b) LPTHE, Universit\'e Pierre et Marie Curie (Paris VI) et Denis
Diderot (Paris VII), Tour 16, 1er. \'etage, 4, Place Jussieu, 
75252 Paris, Cedex 05, France\\
(c) Department of Physics, National Dong Hwa University, Shoufeng, 
Hualien 974, Taiwan, Republic of China\\
(d) Department of Physics and Astronomy, University of North Carolina, 
Chapel Hill, North Carolina 27599}
\date{\today}
\maketitle 

\begin{abstract} 
We study the transport coefficients, damping rates and mean free paths
of soft fermion collective excitations in a hot fermion-gauge-scalar plasma 
with the goal of understanding the main physical mechanisms that determine
transport of chirality in scenarios of non-local electroweak baryogenesis.
The focus is on identifying the {\em different} transport coefficients 
for the different branches of {\em soft} collective excitations of the
fermion spectrum. These branches correspond to collective excitations with
opposite ratios of chirality to helicity and different dispersion relations. 
By combining results from the hard thermal loop (HTL) resummation program
with a novel mechanism of fermion damping through heavy scalar decay, 
we obtain a robust description of the different damping rates and 
mean free paths for the soft collective excitations to leading order 
in HTL and lowest order in the Yukawa coupling. 
The space-time evolution of wave packets of collective excitations
unambiguously reveals the respective mean free paths. 
We find that whereas both the gauge and scalar contribution to the 
damping rates are {\em different} for the different branches, 
the {\em difference} of mean free paths for both branches is mainly
determined by the decay of the heavy scalar into a hard fermion and 
a soft collective excitation. We argue that these mechanisms are 
robust and are therefore relevant for non-local scenarios of baryogenesis
either in the Standard Model or extensions thereof.
\end{abstract} 

\section{Introduction} 

One of the fundamental problems confronting particle astrophysics is 
that of baryogenesis, the origin of the abundance of matter over 
antimatter in the observed Universe. 
The Standard Model (SM)~\cite{kuzmin} and extensions thereof offer the
tantalizing possibility of providing an explanation for baryogenesis with
physics at the electroweak scale (for a description of mechanisms of
baryogenesis see Ref.~\cite{dolgov}). 
Although the Standard Model satisfies
the three criteria for baryogenesis: (i) violation of baryon number, 
(ii) violation of $C$ and $CP$, and (iii) departure from thermal 
equilibrium~\cite{turok,cohen,ruba}, 
a wealth of theoretical evidence combined with experimental bounds on 
the $CP$ violating phases in the Kobayashi-Maskawa (KM) matrix, 
and the current bounds on the Higgs mass, $m_H \geq 88$ GeV, 
from LEP2~\cite{aleph} and the most recent lattice simulations~\cite{csikor} seem to lead to the conclusion that the 
minimal Standard Model~\cite{farrar} may not be able to explain the 
observed baryon asymmetry but perhaps extensions 
could naturally lead to such an explanation 
(for recent reviews see Refs.~\cite{trodden,riotto}). 

An important non-equilibrium ingredient for a successful scenario of
electroweak baryogenesis is a strongly first order phase transition 
that proceeds via the nucleation of the true (broken symmetry) 
vacuum in the background of the false (unbroken symmetry) vacuum 
plus non-trivial topological field configurations, 
the sphalerons, that are responsible for the baryon violating processes~\cite{trodden}. 

Amongst the several proposals for explaining the baryon asymmetry, 
we focus on transport aspects related to non-local or charge 
transfer baryogenesis~\cite{cohen}. This scenario assumes a strongly
first order phase transition in the electroweak theory 
(or extensions thereof)~\cite{trodden,riotto}.
As the nucleated bubbles of the broken phase in the background of 
the unbroken (false vacuum) phase expand, a flux of fermions from the
unbroken phase scatter  
off the bubble walls, $CP$ violating interactions near the wall are
converted to an asymmetry  
in the baryon number via sphaleron processes in the 
unbroken phase~\cite{trodden,farrar}.  

An important ingredient in this scenario is the transport properties 
of chiral fermions in the regime of small spatial 
momentum~\cite{cohen,farrar} but in a {\em plasma}
at temperature $T\approx 100$ GeV. In particular there are two
different regimes  
for non-local baryogenesis depending on whether the mean free path of
the quarks is smaller  
or larger than the width of the bubble walls~\cite{joyce}.

The hot plasma modifies dramatically the properties of
fermions~\cite{klimov,weldon1,pisa},  
in particular when $g\, T \gg  M$ with $g$ the gauge coupling constant and 
$M$ the zero temperature fermion mass, 
the fermion spectrum is characterized by two branches of 
{\em collective excitations}, 
one branch corresponds to the in-medium renormalized fermion, 
with a positive ratio of chirality to helicity, 
and the other branch is a novel collective excitation in the medium, 
the plasmino,  
with a negative ratio of chirality to helicity~\cite{klimov,weldon1}. 
The two branches have a gap $M_{eff} \approx g\,T$ corresponding to 
an effective fermion mass that does not break chiral symmetry~\cite{klimov,weldon1,pisa}. 

The fermion and plasmino branches describe excitations with 
different chirality properties and therefore these are bound to 
play an important role in charge transfer (non-local) mechanisms of
baryogenesis~\cite{trodden,farrar}. 
For soft momenta $k \leq g\,T$ these branches are very different 
from each other, while for large momenta $k \geq g\,T$
they approach the usual fermion dispersion relation, 
and the amplitude of plasmino excitations 
(the wave function renormalization) vanishes exponentially 
fast~\cite{klimov,weldon1,pisa,blaizot1,kapusta,lebellac}. 
Whereas dispersion relations of soft collective excitations in the 
plasma within the context of baryogenesis had been studied 
in Ref.~\cite{farrar} and extended to describe neutralinos
in the medium in supersymmetric extensions in Ref.~\cite{vilja1}, 
the damping or thermalization rate has only been studied at 
zero spatial momentum, i.e., the collective excitations are at rest 
with respect to the plasma~\cite{holo} or for hard spatial 
momentum~\cite{davo} $k \geq T$. 

Early attempts to estimate the diffusion coefficient made use of Boltzmann equations 
and a series of approximations including a simplification of the gluon propagator~\cite{joyce,moore}.
Some of those results were reproduced in~\cite{davo} for quarks of {\em hard} momentum , $k\geq T$.

The two limits that were previously studied, zero spatial momentum and very large spatial 
momentum ($k \geq T$), {\em do not} reveal important
features of the collective excitations which are only present at non-zero but soft momenta 
$0<k\leq g\,T$ where the dispersion relations are
different~\cite{klimov,weldon1,pisa}. At zero spatial momentum, both branches of the fermionic dispersion relation, i.e.,
fermions and plasminos coincide and therefore the damping rates for both collective excitations 
must be the same. In the opposite limit $k \geq T$, the fermion branch approaches the dispersion 
relation of an ordinary vacuum fermion losing all features of the collective behavior, and 
the plasmino branch decouples from the spectrum because its wave function renormalization 
vanishes exponentially. 

In this article we present a detailed study of the relaxation rates (damping or thermalization rates)
and mean free paths for {\em soft} collective excitations with non-vanishing momenta $k \leq g\,T$ 
in a theory that includes  both gauge and Yukawa interactions. 
To our knowledge such study had not been pursued before within the context of gauge and 
scalar (Yukawa) interactions.  Furthermore, since the
different collective excitations have different group velocities, it
is important 
to understand in detail their mean free paths. This
is very relevant for non-local baryogenesis~\cite{cohen,farrar}, 
since the mean free path determines the attenuation length for the transport of chirality by the
different collective excitations. 
As emphasized above, the different branches correspond
to collective excitations with {\em opposite} ratios of chirality to helicity, 
and different dispersion relations.

Our {\em main observations} in this
article are that the transport properties
of these two different collective excitations for soft momenta will in general be {\em different}. 
Since these excitations carry chirality, different damping or relaxation rates will result in 
{\em different mean free paths for collective excitations carrying opposite chirality}. 
As has been argued repeatedly in the literature~\cite{cohen,farrar,trodden} the transport of
chirality plays a fundamental role in non-local baryogenesis, 
thus a consistent microscopic study of baryogenesis in these scenarios must 
necessarily address the transport of chirality by the soft collective excitations and 
is the motivation for our study. 

An important aspect of soft excitations in a plasma is that their consistent description 
requires a non-perturbative resummation scheme, the hard thermal loop (HTL) program~\cite{htl}. 
For soft excitations with $k \leq g\,T$ and vacuum masses $\ll g\,T$
(with $g$ the 
generic gauge coupling constant) the leading corrections from the thermal self-energies 
in the high temperature limit are of ${\cal O}(g\,T)$, hence of the same order as the tree level 
terms in the propagators. 
The leading HTL correction is determined by the hard momentum region in the self-energy 
loop~\cite{weldon1,blaizot1,htl}. However, the damping rate of soft collective excitations 
require that the fermion and gauge bosons propagators and vertices in the self-energy be 
HTL resummed~\cite{brapis,pisarski,rebhan1,iancu}. 
The HTL resummation rendered infrared finite the damping rate of collective excitations 
at rest in the plasma~\cite{brapis}, however the damping rate of {\em moving}
collective excitations is infrared divergent arising from the exchange of transverse gauge 
bosons~\cite{pisarski}. 
Whereas in QCD (and in the SU(2) sector of the electroweak theory) 
the putative non-perturbative magnetic mass provides an
infrared cutoff~\cite{pisarski,rebhan1}, 
in QED (and certainly the U(1) sector of the electroweak theory) such
a possibility  
is unavailable and the infrared divergence remains~\cite{pisarski,rebhan1}. 
The fermion propagator in QED has been  studied within the eikonal
(Bloch-Nordsieck)  approximation in {\em real time}~\cite{iancu}. 
This analysis revealed an anomalous relaxation that explained the nature of
the infrared divergences in the abelian theory and determined that the
relaxation time scale  
of fermionic collective excitations $\propto [g^2 T
\ln 1/g]^{-1}$~\cite{iancu}.  
This anomalous relaxation was recently reproduced via different
methods in scalar QED~\cite{boyrgir}  
that has the same leading HTL structure as QCD and QED for the fermion
and gauge boson propagators 
~\cite{rebhan}. 
Although there is no magnetic screening mass in the abelian theory,
the relaxation scales 
obtained by Refs.~\cite{iancu,boyrgir} are {\em similar} to those obtained in 
QCD~\cite{pisarski,rebhan1}.  

More recently we have obtained the damping rate for fermions in a
fermion-scalar plasma  
and pointed out that the decay of a heavy scalar (Higgs) into fermion pair results 
in a quasiparticle width for the fermion, i.e., the scalar decay leads to a fermion damping 
rate~\cite{fermion}. In this article we combine the results obtained in QCD and QED for the 
damping rate of the collective excitations with the contribution from the scalars to provide a 
detailed understanding of the relaxation scales and mean free paths for the soft chirality 
carrying collective excitations in the plasma. 
 
{\bf The goals:}  In this article we study in detail the damping rates and mean free paths 
of soft collective excitations in a plasma of gauge, fermion and scalar particles as befits 
the scenarios for electroweak baryogenesis. The focus is to provide a quantitative study 
of the {\em different} relaxation scales for the soft collective excitations with opposite 
ratios of chirality to helicity.  

However, rather than studying these aspects in the full Standard Electroweak Model or any 
particular extension of it, we focus on the understanding of the robust and generic features 
of the damping rates and mean free paths of soft collective excitations in an abelian plasma 
of fermions, gauge bosons and scalars with simple Yukawa couplings. 
The main reasons why the results of this simplification {\em are relevant} 
to the physics of the electroweak theory, are that: 

(i) The structure of hard thermal loops is similar in the abelian and non-abelian 
theories~\cite{klimov,weldon1,pisa,blaizot1,lebellac,brapis,pisarski} with
the differences only in the gauge group factors and strength of couplings. 
This is manifestly explicit in similar damping rates obtained for QCD and QED~\cite{pisarski,iancu}. 
A very complete study of the dispersion relations for collective excitations in the electroweak 
plasma has been provided in Ref.~\cite{farrar}. The results of that work clearly show
that the HTL contributions to the self-energies are universal in the sense
that the strong, weak and Higgs sectors yield contributions that are similar
in form and only differ in the appropriate gauge group structure factors. 

(ii) The simple physics that leads to the damping of fermionic excitations via the decay of a heavy
scalar~\cite{fermion} is fairly robust and transcends a particular model, it is mainly described 
by the kinematics of the decay of a heavy scalar into fermion pair in the medium. 

To model the set of parameters of the electroweak theory we assume a scalar
of mass $m \approx 100$ GeV according with the current LEP2 bounds
on the Higgs mass~\cite{aleph}, and a first order phase transition temperature $T\approx 100$ GeV 
which results in $m \approx T$. This will become important in the analysis of the scalar 
contribution to the fermion self-energy. Furthermore we will focus on the lightest quarks 
and leptons for which $g\,T \gg  M$ with $M$ being the fermion vacuum mass. 
Only for the lightest fermions we expect soft collective excitations
in the medium, since for fermions with $M \gg  g\,T$ the in-medium
corrections to the self-energy 
are perturbatively small for weak couplings. For these lightest 
generations, the Yukawa couplings $y \ll  g$ and the scalar contributions to
the fermionic self-energies are perturbative. However, as we will see in
detail below, the scalar contribution to the damping rates and mean free
paths is important.  

{\bf Main Results:} The main results can be summarized with the
statement that for soft collective excitations that carry opposite
ratios of chirality to helicity, i.e., fermions and plasminos, the
damping rates $ \Gamma_{\pm}(k) $ and mean free paths
$\lambda_{\pm}(k)$ are {\em different}.  
The difference in these relaxation scales is maximal in the range of
momenta $0<k\leq g\,T$.   The  difference in the mean free paths of the soft collective
modes is dominated by the scalar
contribution that we explicitly compute in Eq.~(\ref{scalarrate}).
The gauge boson contribution to the damping rate of moving collective
excitations  
arises mainly from the absorption and emission of soft transverse gauge bosons
for which a hard thermal loop resummation of the propagators and vertices
is required. On the other hand the scalar contribution to the damping
rates arises from the {\em decay} of the scalar into a hard single particle
excitation and a soft collective excitation. Both contributions are different
for the different branches, in the gauge case the difference is determined
by the different group velocities. 

We also provide a wave packet analysis to extract the mean free paths
of the moving collective excitations. We find that the {\em difference} of
mean free paths for the collective excitations with different ratios
of chirality to helicity is mainly determined by the {\em scalar}
contribution to the damping rates.  The mean free paths and their
difference are explicitly given by Eqs.~(\ref{meanfre}-\ref{differencemean}).

The article is organized as follows. In Sec.~II we introduce the
renormalized, real-time Dirac equation for fermions in the medium, the
renormalization aspects are important for consistency because we need
to isolate the wave function renormalization factors for the collective
excitations. The main reason for studying the effective in-medium Dirac equation
in {\em real time} is that this formulation allows us to extract unambiguously 
the mean free paths or attenuation lengths by focusing on the time evolution of 
wave packets of collective excitations. This treatment bypasses the assumptions of transport by 
diffusion. Furthermore, in this manner we can make direct
contact with the results of Refs.~\cite{iancu,boyrgir,fermion} and adapt them
to the situation under study. In this section we obtain the
different contributions to the self-energy and renormalize the ultraviolet
divergences. Sec.~III is devoted to a  detailed study of the damping
rates and we establish some of our main results. In Sec.~IV
we study the time evolution of wave packets of collective excitations and extract
the mean free paths or attenuation lengths. We then show that the main
contribution to the {\em difference} between the mean free paths for the
two soft collective excitations arises from the scalar contribution to the damping
rate. Sec.~V summarizes
our results and describes their implications for the standard model 
(or its extensions) and points out the possible modifications to our 
results in extensions of the Standard Model that include light scalars.

\section{Effective Dirac equation in the medium}

An unambiguous identification of the damping rates requires an analysis
of the retarded propagators in the complex frequency plane. The lifetimes
(damping rates) are identified as complex poles in the unphysical Riemann sheet.

Rather than computing the self-energies in imaginary time and
perform the analytic continuations to obtain the retarded self-energies in the complex 
frequency plane, we obtain the effective Dirac equation directly in {\em real time}. This
alternative approach has two main advantages: (i)
it allows us to make contact with the real-time results for the relaxation
of the single particle Green's function presented in
Refs.~\cite{iancu,boyrgir},
(ii) by following the space-time evolution of an initially prepared wave 
packet of collective excitations we can extract directly the mean free paths.

As mentioned in the introduction we consider an abelian theory with one
fermion species and one real scalar coupled to the fermion via a Yukawa
interaction. We will work in the Coulomb gauge in which 
$\nabla\cdot\vec{A}=0$ because
the HTL self-energies have a more clear interpretation in this gauge. 

The self-energies in leading order in HTL have been proven to be the same 
in Coulomb gauge and in covariant and axial gauges~\cite{lebellac,brapis,pisarski}. 
Ref.~\cite{lebellac} provides a discussion of the gauge parameter
independence of the HTL contribution to the self-energies.
 
The Lagrangian density is given by 
\begin{eqnarray}
{\cal L}&=& \bar{\Psi}\left(i{\not\!{\partial}}-g_0 \gamma^0 A^0+ g_0 
\vec{\gamma}\cdot\vec{A}_T-y_0\phi-M_0\right)\Psi+ \bar{\Psi}\eta+\bar{\eta}\Psi 
\nonumber\\
&&+ \frac{1}{2}\left[\left(\partial^\mu \vec{A}_T\right)^2 
+ \left(\nabla A^0\right)^2
+ \left(\partial^\mu\phi\right)^2
- m_0^2 \phi^2\right] + {\cal L}_I[\phi]~,
\end{eqnarray}
where the Grassmann valued source terms were introduced to obtain the
effective Dirac equation in the medium by analyzing the linear response
to these sources~\cite{fermion}. We have not included the possible gauge
couplings of the scalar since these will not contribute to leading order
in the hard thermal loops and lowest order in the Yukawa couplings. 

The self-interaction of the scalar
field accounted for by the term ${\cal L}_I[\phi]$ 
need not be specified to lowest order. 

We now write the bare fields 
$\Psi$, $\phi$, $\vec{A}_T$, $A^0$ and $\eta$ in terms of the renormalized quantities 
(hereafter referred to with a subscript $r$)
by introducing the renormalization constants:
\begin{eqnarray}
&&\Psi=Z_\psi^{1/2}~\Psi_r~,\;\phi=Z_\phi^{1/2}~\phi_r ~,\; \vec{A}_T=Z_{A}^{1/2}\vec{A}_{rT}~,\;
A^0=Z_{A}^{1/2}A^0_r ~,\; \eta=Z_{\psi}^{-1/2}\eta_r~, \nonumber \\
&& y=y_0 Z^{-1}_y Z_\phi^{1/2}Z_\psi ~,\;
g=g_0 Z^{-1}_g Z_A^{1/2} Z_\psi  ~,\label{wfren}\\
&& m^2_0  = \left[\delta_m(T) +m^2(T)\right]Z^{-1}_{\phi}~,\; 
M_0 = \left(\delta_M +M\right)Z^{-1}_{\psi}~.\nonumber 
\end{eqnarray}
With the above definitions, ${\cal L}$ can be expressed as:
\begin{eqnarray}
{\cal L} &=&\bar{\Psi}_r \left(i{\not\!{\partial}}-g\gamma^0 A^0_r 
+ g \vec{\gamma}\cdot\vec{A}_{rT} -y\phi_r - M\right)\Psi_r
+\frac{1}{2}\left[\left(\partial^\mu \vec{A}_{rT}\right)^2 
+\left(\nabla A^0_r\right)^2 +\left(\partial^\mu\phi_r\right)^2 
-m^2(T) \phi_r^2 \right] \nonumber\\
&&+{\cal L}_{rI}[\phi_r] + \bar{\eta}_r\Psi_r
+\bar{\Psi}_r\eta_r + {\cal L}_{\text{c.t.}}~,\label{renlan}
\end{eqnarray}
where $m(T)$, $M$, $g$ and $y$ are the renormalized parameters, and 
the counter-term Lagrangian ${\cal L}_{\text{c.t.}}$ is given by
\begin{eqnarray}
{\cal L}_{\text{c.t.}}&=& \bar{\Psi}_r \left(i\delta_\psi{\not\!{\partial}}-g\delta_g\gamma^0 A^0_r
+ g\delta_g\vec{\gamma}\cdot\vec{A}_{rT} - y \delta_y \phi_r
- \delta_M \right)\Psi_r \nonumber\\
&&+\frac{1}{2}\left[\delta_A\left(\partial^\mu\vec{A}_{rT}\right)^2
+\delta_A\left(\nabla A^0_r \right)^2 
+\delta_\phi \left(\partial^\mu\phi_r\right)^2 
-\delta_m(T) \phi_r^2 \right] 
+\delta{\cal L}_{rI}[\phi_r]~.\label{counterterms}
\end{eqnarray}
The terms with the coefficients
\begin{eqnarray}
&&\delta_\psi=Z_\psi-1~,\;\delta_\phi=Z_\phi-1~,\;
\delta_A=Z_A-1~,\nonumber\\
&&\delta_M= M_0 Z_\psi-M~,\;\delta_m(T)= m^2_0 Z_\phi-m^2(T)~,\\
&&\delta_y=Z_y-1~,\;\delta_g=Z_g-1~,  \nonumber
\end{eqnarray}
and $\delta{\cal L}_{rI}[\phi_r]$ are the counterterms to be 
determined consistently in the perturbative
expansion by choosing a renormalization prescription.  
The main reason for going through the renormalization is to extract
the wave functions corresponding to the collective excitations since these
will be important for the calculation of the damping rates. By choosing
the wave function renormalization $Z_{\psi}$ to be defined on-shell at
zero temperature, we isolate the finite temperature from the zero temperature 
renormalizations unambiguously and recognize the wave function
renormalization for the collective excitations directly.  


We have included the finite temperature corrections to the scalar mass
in the renormalized Lagrangian and explicitly introduced a finite
temperature counterterm. The scalar mass receives finite temperature
corrections also from hard thermal loops, proportional to $T^2$. 
As it will become clear below we are interested in the kinematic decay of
the Higgs into the fermionic collective modes which will contribute to
their mean free paths. By including
the finite temperature corrected Higgs  mass in the Lagrangian allows to
treat the kinematic decay in terms of the in-medium corrected mass directly.
The counterterm $\delta_m(T)$ will be chosen consistently to cancel the
finite temperature corrections to the mass. There is a small caveat, 
however, we envisage our analysis to be valid both in the (homogeneous) unbroken and the
broken symmetry phases in the case of a strongly first order phase
transition. The mass term in the Lagrangian should then be defined
as the position of the pole of the finite temperature propagator for the Higgs. We will implicitly use 
this definition of the Higgs mass in what follows.  Furthermore in our analysis
we will {\em assume} that in either case, unbroken or broken symmetry phases
the effective Higgs mass is large enough that its kinematic decay into
the soft fermionic collective excitations is allowed, in particular we
{\em assume} that $m(T)\gg g\,T$. In the broken symmetry phase and for a strong
first order phase transition $m^2(T) \approx m^2(0)$ with 
$m(0) \approx T \approx 100~\mbox{Gev}$ this assumption seems reasonable, 
in the unbroken phase whether this condition is satisfied depends on the 
details and strength of the first order phase transition.

 If the effective mass in
either phase does not allow the kinematical decay of the in-medium
Higgs into soft fermionic collective modes then there is no 
Higgs contribution to their mean free path. 

Estimates based on the finite temperature effective potential would seem to suggest that the Higgs mass may not
be much heavier than $gT_c$\cite{wagner,quiros}, but conclusions based on effective potentials in gauge theories are 
fraught with ambiguities such as the gauge dependence of the effective potential which translates into gauge 
dependence of off-shell quantities, and the fact that the effective potential becomes complex\cite{meta,wu,boyaveff} near
its minimum. A Higgs mass defined as the second derivative of the effective potential at its minimum is not an
on-shell quantity\cite{boyaveff} (since the effective potential is defined at zero four momentum), only the position of the pole
of the Higgs propagator is an on-shell and therefore gauge invariant quantity.  Given the uncertainties on the
scalar sector of the Standard Model, only a sound {\em experimental} bound can validate or rule out the assumption  used in this work on the value of the
Higgs mass. 


We must also point out that our analysis is valid  in the unbroken or
the broken symmetry phases (under the proviso of the kinematic constraint
for the decay of the Higgs) but certainly {\em not} near the wall of a
nucleating bubble. The nature of the soft fermionic collective excitations
in the presence of a wall is not well understood, since the hard thermal
loop resummation that leads to the collective excitations is only valid
in an homogeneous equilibrium background. Thus before we can extrapolate
our analysis to the case of collective fermionic excitations near a bubble
wall a consistent study of the effects of the wall on these collective 
modes must be pursued. As mentioned above our goal is to provide a quantitative
analysis of the mean free paths of the soft fermionic collective modes at
least in the homogeneous phases, with particular attention to the {\em 
difference} of mean free paths of the different collective modes.  


The dynamics of expectation values and correlation functions of the
quantum field is obtained by implementing the Schwinger-Keldysh 
closed-time-path formulation of non-equilibrium 
quantum field theory~\cite{ctp,disip,linon}.  
The main ingredient in this formulation is the {\em real time} 
evolution of an initially prepared density matrix and its 
path integral representation. 
It requires a path integral defined along a closed
time path contour.  This formulation has been
described elsewhere within many different contexts and we refer the 
reader to the literature for details~\cite{ctp,disip,linon}. 

The effective Dirac equation is the equation of motion for the expectation
value of the fermion field 
$\psi(\vec{x},t) \equiv \langle \Psi_r(\vec{x},t)\rangle_{\eta}$ 
in the presence of the sources $\bar{\eta},\eta$. 
Since the details and methods to obtain the effective
Dirac equation in the medium have been described at length
elsewhere~\cite{fermion},  
we refer the reader to Refs.~\cite{fermion,disip,linon}
and simply quote the final result: 
\begin{equation}
\left[\left(i\gamma_0\frac{\partial}{\partial t} 
- \vec{\gamma}\cdot\vec{k}-M \right)+
\delta_\psi\left(i\gamma_0\frac{\partial}{\partial t}
- \vec{\gamma}\cdot\vec{k}\right)
-\delta_M\right] \psi_{\vec{k}}(t)+
\int_{-\infty}^t dt'~  \Sigma_{\vec{k}}(t-t')\; \psi_{\vec{k}}(t')= -
\eta_{r\vec{k}}(t)~, \label{diraceqn}
\end{equation}
where
$$
\psi_{\vec{k}}(t) \equiv \int d^3x\;  e^{-i \vec{k} \cdot \vec{x} } \;
\psi(\vec{x},t) \;\;.
$$
and $\Sigma_{\vec k}(t-t')$ is the
total self-energy, which to lowest order is the sum of the transverse, 
longitudinal and Yukawa contributions
respectively, i.e., 
$\Sigma_{\vec k}(t-t')=\Sigma^T_{\vec k}(t-t')+\Sigma^L_{\vec
k}(t-t')+\Sigma^y_{\vec k}(t-t')$.  
The scalar contribution can be found in~\cite{fermion}, and for the gauge contribution
to the fermion self-energy the main ingredients are the real-time propagators
\begin{eqnarray}
&&{\langle}{A}^{(a)}_{rTi}(\vec{x},t){A}^{(b)}_{rTj}(\vec{x}',
t^{\prime}){\rangle}=-i\int {d^3k\over{(2\pi)^3}}\;{\cal G}_{ij}^{ab}
(\vec{k};t,t^\prime)\;e^{i\vec{k}\cdot(\vec{x}-\vec{x}')}\;,\nonumber\\
&&{\cal G}_{ij}^{++}(\vec{k};t,t^\prime)={\cal P}_{ij}(\vec{k})
\left[{\cal G}_{k}^{>}(t,t^{\prime})\Theta(t-t^{\prime})
+{\cal G}_{k}^{<}(t,t^{\prime})\Theta(t^{\prime}-t)
\right]\;,\nonumber\\
&&{\cal G}_{ij}^{--}(\vec{k};t,t^\prime)={\cal P}_{ij}(\vec{k})
\left[{\cal G}_{k}^{>}(t,t^{\prime})\Theta(t^{\prime}-t)
+{\cal G}_{k}^{<}(t,t^{\prime})\Theta(t-t^{\prime})
\right]\;,\nonumber\\
&&{\cal G}_{ij}^{+-}(\vec{k};t,t^\prime)={\cal P}_{ij}(\vec{k})
{\cal G}_{k}^{<}(t,t^{\prime})\;,\nonumber\\
&&{\cal G}_{ij}^{-+}(\vec{k};t,t^\prime)={\cal P}_{ij}(\vec{k})
{\cal G}_{k}^{>}(t,t^{\prime})\;,\nonumber\\
&&{\cal G}_{k}^{>}(t,t^{\prime})=
\frac{i}{2k}\left[(1+\tilde{n}_k)e^{-ik(t-t^\prime)}
+\tilde{n}_k e^{ik(t-t^\prime)}\right]\;,\nonumber\\
&&{\cal G}_{k}^{<}(t,t^{\prime})=\frac{i}{2k}\left[
\tilde{n}_k e^{-ik(t-t^\prime)}
+(1+\tilde{n}_k)e^{ik(t-t^\prime)}\right],\nonumber\\
&&\tilde{n}_k=\frac{1}{e^{\beta k}-1}\;,\quad \beta=\frac{1}{T}\;,\quad
{\cal P}_{ij}(\vec{k})=\delta_{ij}-\frac{k_ik_j}{k^2}\;.\label{projector}
\end{eqnarray}

The retarded time evolution must be understood as an initial
value problem~\cite{fermion,disip,linon}, which is set up consistently
by taking the  source  to be switched on
adiabatically from $t=-\infty$ and 
switched off at $t=0$ thus providing the following initial conditions
to the mean field 
\begin{eqnarray}
&&\psi_{\vec{k}}(t=0)=\psi_{\vec{k}}(0)~,\quad\dot{\psi}_{\vec{k}}(t <
0)=0~. \label{iniconds} 
\end{eqnarray}

It is convenient to separate the memory integral in (\ref{diraceqn}) from 
$t'=-\infty $ to $t'=0$  and $t'=0$ to 
$t'=t$, and for this purpose  we introduce the kernel $\sigma_{\vec{k}}(t-t')$ 
defined by the relation
\begin{equation}
\frac{d}{dt'}
\sigma_{\vec{k}}(t-t')= \Sigma_{\vec{k}}(t-t')~.
\label{sigma}
\end{equation}
\noindent Since $\eta_{r\vec{k}}(t>0)=0$, we integrate by parts the
memory kernel and finally obtain the equation of motion for $t>0$ in
the following form 
\begin{eqnarray}
&&\left[\left(i\gamma_0\frac{\partial}{\partial t}
- \vec{\gamma}\cdot\vec{k}-M\right)+
\delta_\psi\left(i\gamma_0\frac{\partial}{\partial t}
- \vec{\gamma}\cdot\vec{k}\right)
+ \sigma_{\vec{k}}(0)-\delta_M \right] 
\psi_{\vec{k}}(t) -
\int_0^t dt'\; \sigma_{\vec{k}}(t-t')\;
\dot{\psi}_{\vec{k}}(t')=0~\;.\label{eom}
\end{eqnarray}
The equation of motion (\ref{eom}) can now be solved by Laplace transform
as befits an initial value problem.
The Laplace transformed equation of motion is given by
\begin{eqnarray}
&&\left[i\gamma_0 s
- \vec{\gamma}\cdot\vec{k}-M+ \delta_\psi 
\left(i \gamma_0 s -\vec{\gamma}\cdot\vec{k}\right)-\delta_M 
+ \sigma_{\vec{k}}(0)-
s\tilde{\sigma}_{\vec{k}}(s) \right] \tilde{\psi}_{\vec{k}}(s)= 
\left[i\gamma_0 + i\delta_\psi \gamma_0- 
\tilde{\sigma}_{\vec{k}}(s)\right]\psi_{\vec{k}}(0)~, \label{lapladiraceqn}
\end{eqnarray}
where $\tilde{\psi}_{\vec{k}}(s)$ and $\tilde{\sigma}_{\vec{k}}(s)$ 
are the Laplace transforms of $\psi_{\vec{k}}(t)$ and
$\sigma_{\vec{k}}(t)$ respectively:
$$
\tilde{\psi}_{\vec{k}}(s)\equiv \int_0^{\infty} dt \; e^{-st}\;
\psi_{\vec{k}}(t) \quad , \quad \tilde{\sigma}_{\vec{k}}(s)\equiv
\int_0^{\infty} dt \; e^{-st}\; \sigma_{\vec{k}}(t)\;.
$$
Renormalization proceeds by requiring that the counterterms cancel the divergent parts 
of the self-energy. 

Specifically, the renormalized self-energy contributions are given by
\begin{eqnarray}
\tilde{\Sigma}_{r\vec k}(s) & = &\sigma_{r\vec k}(0)-s~\tilde{\sigma}_{r\vec k}(s)~, \nonumber \\
\sigma_{r\vec k}(0) & = &  \sigma_{\vec k}(0)-\delta M - \delta_{\psi}\vec{\gamma}\cdot {\vec k}~, 
\quad\tilde{\sigma}_{r\vec k}(s)  =  \tilde{\sigma}_{r\vec k}(s)-i\gamma_0 \delta_{\psi}~.
\label{renorselfener}
\end{eqnarray}

The counterterms will be displayed explicitly below. 
We thus obtain the fully renormalized  effective Dirac equation in the medium, which is given by
\begin{eqnarray}
&&\left[i\gamma_0 s
- \vec{\gamma}\cdot\vec{k}-M + \tilde{\Sigma}_{r\vec{k}}(s)\right]
\tilde{\psi}_{\vec{k}}(s)= 
\left[i\gamma_0 - 
\tilde{\sigma}_{r\vec{k}}(s)\right]\psi_{\vec{k}}(0)~\;,\label{reneom}
\end{eqnarray}

The solution to (\ref{reneom}) is therefore given by 
\begin{equation}
\tilde{\psi}_{\vec{k}}(s)= \frac{1}{s}\left[ 1 + S(s,\vec{k}) 
\left(\vec{\gamma}\cdot\vec{k}+M -\tilde{\Sigma}_{r\vec{k}}(0)\right)\right]
\psi_{\vec k}(0)~,
\label{laplakern}
\end{equation}
where we have introduced the full renormalized fermion propagator in
terms of the Laplace  
variable $s$ 
\begin{equation}
S(s,\vec{k})=\left[i\gamma_0 s - \vec{\gamma}\cdot\vec{k}-M
+\tilde{\Sigma}_{r\vec{k}}(s)\right]^{-1}~.
\end{equation} 
The retarded propagator is obtained via the analytic continuation
$s=-i\omega + 0^+$.  
Stable excitations
correspond to isolated poles of $S(s=-i\omega + 0^+,k)$ in the
physical Riemann sheet  
in the complex $\omega$ plane,
whereas resonances correspond to complex poles in the second (or
higher) Riemann sheet.  
These resonances are
quasiparticles with lifetimes determined by the imaginary part of the
complex pole and  
the spectral density for the fermion propagator features a
Breit-Wigner form with a narrow peak  
in the weak coupling limit.

\subsection{The self-energy}
Our main goal is to provide a description of damping and transport of soft fermionic 
collective excitations with typical momenta
$k \leq g\,T$.  In this region of soft momenta, the hard thermal loop contribution from 
gauge bosons must be treated non-perturbatively since it is of order $g\,T$. 
In principle if there are HTL contribution 
from the scalars, it should also be treated non-perturbatively. 
However we focus our study on the light quarks and leptons 
for which the typical Yukawa couplings $y\ll g$, the HTL 
contribution from the gauge bosons gives the only non-perturbative contribution 
(at least to lowest order).

Using the non-equilibrium Green's functions for free fields given in Ref.~\cite{fermion} 
and Eqs.~(\ref{projector}), we find to one loop order the following contributions 
to the fermion self-energy: 
\begin{eqnarray}
&&\Sigma^{y}_{\vec{k}}(t-t')=i\gamma_0\; 
\Sigma^{y(0)}_{\vec{k}}(t-t')+
\vec{\gamma}\cdot\vec{k}\; 
\Sigma^{y(1)}_{\vec{k}}(t-t')
+\Sigma^{y(2)}_{\vec{k}}(t-t')~, \\
&&\Sigma^{T}_{\vec{k}}(t-t')=i\gamma_0\; 
\Sigma^{T(0)}_{\vec{k}}(t-t')+
\vec{\gamma}\cdot\vec{k}\; 
\Sigma^{T(1)}_{\vec{k}}(t-t')
+\Sigma^{T(2)}_{\vec{k}}(t-t')~, \\
&&\Sigma^{L}_{\vec{k}}(t-t')=g^2\int\frac{d^3 q}{(2\pi)^3}
\frac{1-2\bar{n}_q}{2\bar{\omega}_q 
(\vec{k}+\vec{q})^2}(\vec{\gamma}\cdot\vec{q}-M)~\delta(t-t')~,
\end{eqnarray}
where $\Sigma^{y(i)}_{\vec{k}}(t-t')$ to one loop order ${\cal O}(y^2)$ are given in 
Ref.~\cite{fermion}. Although the
transverse  self-energy is available in the literature in the 
{\em imaginary time} formulation
of finite temperature field theory, their form in {\em real time}  is not readily available. 
It is given by
\begin{eqnarray}
\Sigma^{T(0)}_{\vec{k}}(t-t')
&=& g^2 \int \frac{d^3 q}{(2\pi)^3 \tilde{\omega}_{k+q}}\nonumber\\
&&\times
\Bigl[\cos[(\tilde{\omega}_{k+q}+\bar{\omega}_q)(t-t')]
(1+\tilde{n}_{k+q}-\bar{n}_q)+
\cos[(\tilde{\omega}_{k+q}-\bar{\omega}_q)(t-t')]
(\tilde{n}_{k+q}+\bar{n}_q)\Bigr]~,\label{sigmaT0}\\
\Sigma^{T(1)}_{\vec{k}}(t-t')
&=& g^2 \int \frac{d^3 q}{(2\pi)^3 \bar{\omega}_{q}\tilde{\omega}_{k+q}} 
\left[\frac{\vec{k}\cdot(\vec{k}+\vec{q})\;\vec{q}
\cdot(\vec{k}+\vec{q})}{k^2(\vec{k}+\vec{q})^2}\right]\nonumber\\ 
&&\times
\Bigl[\sin[(\tilde{\omega}_{k+q}+\bar{\omega}_q)(t-t')]
(1+\tilde{n}_{k+q}-\bar{n}_q)-
\sin[(\tilde{\omega}_{k+q}-\bar{\omega}_q)(t-t')]
(\tilde{n}_{k+q}+\bar{n}_q)\Bigr]~,\label{sigmaT1}\\
\Sigma^{T(2)}_{\vec{k}}(t-t')
&=& -g^2 M \int \frac{d^3 q}{(2\pi)^3 \bar{\omega}_{q}\tilde{\omega}_{k+q}}\nonumber\\
&&\times
\Bigl[\sin[(\tilde{\omega}_{k+q}+\bar{\omega}_q)(t-t')]
(1+\tilde{n}_{k+q}-\bar{n}_q)-
\sin[(\tilde{\omega}_{k+q}-\bar{\omega}_q)(t-t')]
(\tilde{n}_{k+q}+\bar{n}_q)\Bigr]~,\label{sigmaT2}
\end{eqnarray}
with $\bar{\omega}_q=\sqrt{M^2+\vec{q}^2}$, $\tilde{\omega}_{k+q}=\sqrt{(\vec{k}+\vec{q})^2}$, 
$\bar{n}_q=(e^{\beta\bar{\omega}_q}+1)^{-1}$ 
and $\tilde{n}_{k+q}=(e^{\beta\tilde{\omega}_{k+q}}-1)^{-1}$ being
the respective energies and distribution functions for the fermion and the gauge boson in the loop.

These contributions to the self-energy are solely in terms of the bare propagators 
for the internal lines in the loop.
For the gauge contribution to the self-energy this is only valid for  loop momenta  
$k\gg g\,T$, this is the region of the
loop integral that gives rise to the HTL non-perturbative part. The region of soft 
loop momenta $k \leq g\,T$ requires the HTL 
dressing of the internal propagators and vertices~\cite{brapis,pisarski,iancu} and  
will be discussed in detail later. 

For the scalar contribution in principle one should consider the different regions of 
the loop momentum, however, since the scalar is taken to represent a Higgs of mass 
$\approx 90-100~\mbox{GeV} \approx T$
the scalar line does not require HTL resummation. However as argued above, the scalar 
self-energy is perturbatively
small as compared to the HTL part of the gauge boson self-energy. 
Its real part  will yield a perturbatively small
correction to the dispersion relation of the collective excitations and will be neglected, 
however the {\em imaginary} part will lead to
a contribution to the damping rate, which is important since in lowest order HTL the collective 
excitations are stable. We anticipate now and it will be argued more forcefully 
below, that for a Higgs
mass of ${\cal O}(T)$ the energy conservation constraint for the imaginary part requires 
that for soft external momentum
of the fermion, the {\em internal} fermion line must carry a momentum at least  
of ${\cal O}(T)$ and therefore in order
to obtain the scalar contribution to the damping rate the internal fermion line 
{\em does not} need HTL resummation.
This is a very important point to which we will come back below and it results 
in that the scalar contribution
does not suffer from the infrared sensitivity associated with the gauge boson 
contribution to the self-energy.  

Before we discuss each contribution to the self-energy in detail, we proceed 
to fix the counterterms to render the effective Dirac equation finite to one loop order. 

The ultraviolet-divergent parts of $\sigma_{\vec{k}}(0)$ and
$\tilde{\sigma}_{\vec{k}}(s)$ entering in the Dirac equation (\ref{lapladiraceqn}) 
are contained in the zero temperature 
contributions. We use dimensional regularization in three spatial 
dimensions, introduce a renormalization scale ${\cal K}$ and find 
\begin{eqnarray}
\sigma_{\vec{k}}(0)&=&
-\left[\left(\frac{y^2}{16\pi^2}+\frac{g^2}{8\pi^2}\right)
\vec{\gamma}\cdot\vec{k}
-\left(\frac{y^2}{8\pi^2}-\frac{g^2}{2\pi^2}\right)M\right]
\frac{{\cal K}^{-\epsilon}}{\epsilon}
+\mbox{finite}~,\nonumber\\
\tilde{\sigma}_{\vec{k}}(s)&=&
-i\gamma_0\left(\frac{y^2}{16\pi^2}+\frac{g^2}{8\pi^2}\right)
\frac{{\cal K}^{-\epsilon}}{\epsilon}+\mbox{finite}~,
\end{eqnarray}
where the finite parts contain the finite temperature contributions plus
zero temperature finite terms. The counterterms can then be chosen in minimal subtraction 
to be given by
\begin{equation}
\delta_{\psi}=-\left(\frac{y^2}{16\pi^2}+\frac{g^2}{8\pi^2}\right)
\frac{{\cal K}^{-\epsilon}}{\epsilon}~,\quad
\delta_{M}=M\left(\frac{y^2}{8\pi^2}-\frac{g^2}{2\pi^2}\right)
\frac{{\cal K}^{-\epsilon}}{\epsilon}~.
\end{equation}
A convenient choice corresponds to on-shell renormalization at zero
temperature, since in this case the residue at the poles of the
propagators yield directly the wave function renormalizations of the
collective excitations.  

\section{The In-medium Fermion Propagator}

After renormalization the zero temperature part of the self-energy is
finite and perturbatively small, therefore we can neglect it and focus
on the finite temperature part. 

As mentioned in the introduction, our goal is to describe the transport of chirality 
by the soft collective excitations. Since the typical size of a non-perturbative gauge 
field configuration
is ${\cal O}(1/g^2T)$~\cite{trodden}, the contributions to baryon violating processes 
from quarks and leptons with 
soft momenta $\leq g\,T$ is important~\cite{farrar}. For these soft fermionic excitations, the 
HTL contribution to the self-energy
given by the hard momentum region of the loop integral (i.e., $k \geq T$) is 
${\cal O}(g\,T)$ and must be treated non-perturbatively.

Furthermore in the region near the electroweak phase transition with $T \approx 100$ GeV 
and $g \approx 0.3-0.6$ one finds that $g\,T \gg  M$ and the vacuum mass term can be 
neglected for the lightest leptons and quarks.

Therefore: (i) we neglect the zero temperature
contribution, (ii) neglect the (renormalized) fermion mass, (iii) separate
the HTL non-perturbative contribution from the hard gauge boson exchange from 
the perturbative contribution of soft gauge boson exchange and scalar
exchange. Thus the  gauge contribution to the fermion self-energy is separated 
into the hard thermal loop part $\delta\tilde{\Sigma}_{r\vec{k}}(s)$ 
which is given by the hard loop momentum region $k \approx T$, and a correction
$ {}^\ast \tilde{\Sigma}_{r\vec{k}}(s) $~\cite{pisarski} which arises
from the soft  region of the loop integration $k \leq g\,T$. In this
region  the internal fermions and gauge bosons as well as the vertices
must be HTL resummed and a detailed analysis finds  that this
contribution is of order $ g^2T $~\cite{pisarski}  
(see below). Whereas the HTL contribution only gives an imaginary part
below the light cone (Landau damping), the $g^2 T$ correction gives a
contribution to the damping rate~\cite{pisarski} both at rest and for a
moving fermion. Since the Yukawa coupling for the lightest quarks and
leptons is $ y\approx 10^{-4} \ll  g $ and the ratio of the scales that 
enter in the scalar loop is $m/T \approx 1$, the scalar contribution will
be treated perturbatively, along with that of ${}^\ast
\tilde{\Sigma}_{r\vec{k}}(s)$.

Thus to this order, the renormalized full inverse fermion propagator is 
\begin{equation}
S^{-1}(s,\vec{k})=i\gamma_0 s - \vec{\gamma}\cdot\vec{k}
+\delta\tilde{\Sigma}_{r\vec{k}}(s)
+{}^\ast \tilde{\Sigma}_{r\vec{k}}(s)
+\tilde{\Sigma}^y_{r\vec{k}}(s)~.
\end{equation} 
It is convenient to write the full fermion propagator
in terms of a non-perturbative and a perturbative part
\begin{eqnarray}
S^{-1}(s,\vec{k}) & = & S^{-1}_{NP}(s,\vec{k})+S^{-1}_P(s,\vec{k})~, \nonumber \\
S^{-1}_{NP}(s,\vec{k}) & = & i\gamma_0 s - \vec{\gamma}\cdot\vec{k}
+\delta\tilde{\Sigma}_{r\vec{k}}(s)~, \nonumber \\
S^{-1}_P(s,\vec{k}) & = & {}^\ast \tilde{\Sigma}_{r\vec{k}}(s)
+\tilde{\Sigma}^y_{r\vec{k}}(s)~. \label{propagatorsplit}
\end{eqnarray}

The non-perturbative term will determine the position of the poles, i.e.,
the dispersion relation of the
quasiparticles. In the HTL approximation it is known that the poles correspond to 
stable collective excitations, because
the imaginary part of the self-energy is non-vanishing only below the light cone 
(Landau damping). Therefore the damping rate for on-shell excitations vanishes to 
leading order in HTL. The real part of the perturbative contribution will provide a small 
shift to the dispersion relation, whereas
the imaginary part will provide the damping rate of the fermionic collective excitations 
on-shell. Because the correction
to the dispersion relations from the perturbative part is small, 
we will neglect it but instead focus on the
imaginary part since it will determine the damping rate which vanishes 
to leading order in the HTL. 
 
Since the fermion self-energy is a sum of the contribution of the gauge bosons and 
that of the scalars, we study each contribution separately. 

\subsection{Non-perturbative contribution to the fermion self-energy: hard thermal 
loops and collective excitations}

We now focus on $S^{-1}_{NP}(s,\vec{k})$, which is given by the bare propagator plus the HTL 
contribution from the
gauge boson exchange. 
The HTL approximation  is obtained by considering the region $q\approx T$ in the 
loop integrals, i.e., the loop momentum is hard and the external momentum is soft $k \leq g\,T$. 

The HTL contribution to the fermion self-energy is obtained directly 
from the Laplace transform of Eqs.~(\ref{sigmaT0}-\ref{sigmaT2}) in the HTL limit. 
In terms of the Laplace variable $s$, it is found to be given by
\begin{equation}
\delta\tilde{\Sigma}_{r\vec{k}}(s)=
-\frac{g^2 T^2}{16k}\left\{\gamma_0\ln\left(\frac{is+k}{is-k}\right)
+\vec{\gamma}\cdot\hat{k}
\left[2-\frac{is}{k}\ln\left(\frac{is+k}{is-k}\right)\right]\right\}~.
\end{equation}

The analytic continuation of $\delta\tilde{\Sigma}_{r\vec{k}}(s)$
in the complex $s$-plane reads
\begin{eqnarray}
&&\delta\tilde{\Sigma}_{r\vec{k}}(s=-i\omega\pm 0^+)\equiv
\delta{\Sigma}_{\vec{k}}(\omega)=
\delta{\Sigma}_{R\vec{k}}(\omega)\pm i
\delta{\Sigma}_{I\vec{k}}(\omega)~,
\end{eqnarray}
where 
\begin{eqnarray}
\delta{\Sigma}_{R\vec{k}}(\omega)&=&
-\frac{g^2 T^2}{16k}\left\{\gamma_0\ln\left|
\frac{\omega+k}{\omega-k}\right|+\vec{\gamma}\cdot\hat{k}
\left[2-\frac{\omega}{k}\ln\left|\frac{\omega+k}{\omega-k}
\right|\right]\right\}~,\\
\delta{\Sigma}_{I\vec{k}}(\omega)&=&
\frac{\pi g^2 T^2}{16k}\left[\gamma_0-
\vec{\gamma}\cdot\hat{k}\, \frac{\omega}{k}\right]
\Theta(k^2-\omega^2)~,
\end{eqnarray}
and $\hat{k}=\vec{k}/k$.

To leading order in $g$ (HTL), the full fermion propagator is given by
${}^{\ast}S^{-1}(\omega,\vec{k})=\omega\gamma_0-\vec{\gamma}\cdot\vec{k}
+\delta{\Sigma}_{\vec{k}}(\omega)$ which can be written in the form~\cite{weldon1,lebellac}
\begin{equation}
{}^{\ast}S(\omega,\vec{k})=
\frac{1}{2}\left[{}^{\ast}\Delta_{+}(\omega,\vec{k})(\gamma_0-
\vec{\gamma}\cdot\hat{k})+{}^{\ast}\Delta_{-}(\omega,\vec{k})
(\gamma_0+\vec{\gamma}\cdot\hat{k})\right]~,
\end{equation}
where
\begin{eqnarray}
&&{}^{\ast}\Delta^{-1}_{\pm}(\omega,\vec{k})=
\omega \mp k - \frac{M^2_{eff}}{k} 
\left[\left(1\mp \frac{\omega}{k}\right)
Q_{0}\left(\frac{\omega}{k}\right)\pm 1\right]~,\label{deltast}\\
&&Q_{0}\left(\frac{\omega}{k}\right)=\frac{1}{2}\left[
\ln \left|\frac{\omega+k}{\omega-k}\right|- i \pi \Theta(k^2-\omega^2)\right]~.
\end{eqnarray}
Here, $M_{eff}=g T/\sqrt{8}$ is the
effective fermion thermal mass induced by the gauge coupling.

The collective excitations correspond to the poles of the fermion
propagator: there are  
two branches for positive energy (and two for negative energy)
which are the solutions of ${}^{\ast}\Delta^{-1}_{\pm}(\omega,\vec{k})=0$, 
the Dirac spinors associated with these solutions have opposite ratios
of chirality 
to helicity $\chi$ and the residues at the poles are $ Z_{\pm}(k) $,
respectively, with~\cite{klimov,weldon1,pisa,blaizot1,kapusta,lebellac}
\begin{equation}  
 {}^{\ast}\Delta^{-1}_{\pm}(\omega,\vec{k})=0 \Longrightarrow 
 \omega=\omega_{\pm}(k)~,
 \quad Z_{\pm}(k) = \frac{1}{2 M^2_{eff}} \left[\omega^2_{\pm}(k)-k^2\right]~,
 \label{HTLpoles} 
\end{equation}
where the dispersion relations $\omega_{\pm}(k)$ are shown in Fig.~1. 
The upper branch corresponds to $\omega_{+}(k)$ and 
describes collective excitations with ratio of chirality to helicity
$\chi=+1$; the lower branch corresponds to $\omega_{-}(k)$ describing
collective excitations with ratio of chirality to helicity $\chi=-1$. 
The upper branch corresponds to the usual 
fermionic excitation, whereas the lower branch has vanishing group velocity 
(the minimum in the dispersion relation) at $k_{min} \approx 0.4\; M_{eff}$ and describes 
a new collective excitation in the medium, it has been named the 
plasmino to emphasize that it is a fermionic excitation that only exists as a 
collective excitation in the plasma~\cite{klimov,weldon1,pisa,lebellac}. 
The vanishing group velocity at $k_{min}$ is a novel feature of 
the plasmino collective excitation and it will be important for the damping rate
and mean free path.  

We now collect some of the relevant properties of these solutions which
will be relevant for the interpretation of the results (see Ref.~\cite{lebellac} for details)
\begin{eqnarray}
k \ll  M_{eff} & : &  \omega_{\pm}(k) \approx M_{eff}\pm
\frac{k}{3}~,\quad Z_{\pm}(k)  
\approx \frac{1}{2} \pm \frac{k}{3M_{eff}}~, \nonumber \\
k \gg  M_{eff} & : &  \omega_{+}(k) \approx k+ \frac{M^2_{eff}}{k}~,
\quad Z_{+}(k)  
\approx 1+
\frac{M^2_{eff}}{2k^2}\left(1-\ln\frac{2k^2}{M^2_{eff}}\right)~, 
\nonumber \\
&& \omega_{-}(k) \approx k+
\frac{2k}{g}\exp\left(-\frac{2k^2}{M^2_{eff}}\right)~, 
 \quad Z_{-}(k)  \approx \frac{2k^2}{g
M^2_{eff}}\exp\left(-\frac{2k^2}{M^2_{eff}}\right)~. 
\label{properties}
\end{eqnarray}

Two noteworthy aspects of these expressions will be important for the interpretation 
of particular features of
the damping rates: 
(i) the dispersion relations are always above the light cone, (ii) the wave function 
renormalization
(residue at the pole) $Z_-(k)$ vanishes very fast for $k \geq k_{min}$. Thus the collective 
excitation with $\chi = -1$ only contributes in the region of very soft external momentum 
$k \leq k_{min}$.

\subsection{Perturbative corrections: the damping rates} 

Neglecting the fermion vacuum mass term, the general form of the
self-energies is dictated by rotational symmetry, hence
${}^\ast\tilde{\Sigma}_{r\vec{k}}(s)$ and  $\tilde{\Sigma}^y_{r\vec
k}(s)$ can be written as 
\begin{eqnarray}
{}^\ast\tilde{\Sigma}_{r\vec{k}}(s) & = &
-is\gamma_0\,{}^\ast\tilde{\Sigma}^{(0)}_{\vec{k}} 
(s)+\vec{\gamma}\cdot\vec{k} \; {}^\ast\tilde{\Sigma}^{(1)}_{\vec{k}}(s)~,
\label{gaugepert} \\ 
\tilde{\Sigma}^y_{r\vec{k}}(s) & = & 
i\gamma_0 \; s \; \tilde{\varepsilon}^{(0)}_{\vec{k}} (s)+
\vec{\gamma}\cdot
\vec{k} \;\tilde{\varepsilon}^{(1)}_{\vec{k}}(s)~,\label{epsilons2} 
\end{eqnarray}
where we have used the notation of Ref.~\cite{fermion} for
$\tilde{\Sigma}^y_{r\vec k}(s)$. The analytic continuation of these
contributions in the complex $s$-plane are defined by
\begin{eqnarray}
{}^\ast{\Sigma}_{r\vec{k}}(\omega) & \equiv & 
{}^\ast\tilde{\Sigma}_{r\vec{k}}(s=-i\omega\pm 0^+)=
{}^\ast{\Sigma}_{R\vec{k}}(\omega)\pm i \;
{}^\ast{\Sigma}_{I\vec{k}}(\omega)~. \label{imaggauge} \\
{\Sigma}^y_{r\vec{k}}(\omega) & \equiv & 
\tilde{\Sigma}^y_{r\vec{k}}(s=-i\omega\pm 0^+)=
{\Sigma}^y_{R\vec{k}}(\omega)\pm i\;
{\Sigma}^y_{I\vec{k}}(\omega)~. \label{imagscalar} 
\end{eqnarray}

To order $g^2$ and $y^2$, the full fermion propagator now reads
\begin{equation}
S(\omega,\vec{k})=
\frac{1}{2}\left[\overline{\Delta}_{+}(\omega,\vec{k})(\gamma_0-
\vec{\gamma}\cdot\hat{k})+\overline{\Delta}_{-}(\omega,\vec{k})
(\gamma_0+\vec{\gamma}\cdot\hat{k})\right]~,
\end{equation}
where
\begin{eqnarray}
&&\overline{\Delta}^{-1}_{\pm}(\omega,\vec{k})=
{}^\ast\Delta^{-1}_{\pm}(\omega,\vec{k})
+\Pi_{\pm}(\omega,\vec{k})~,\\
&&\Pi_{\pm}(\omega,\vec{k})=
\frac{1}{4}
\text{Tr}\left[(\gamma_0\mp\vec{\gamma}\cdot\hat{k})\,\left({}^\ast\Sigma_{r\vec{k}}(\omega)+
\Sigma^y_{r\vec k}(\omega) \right)
\right]~. \label{totalpi}
\end{eqnarray}
with ${}^{\ast}\Delta^{-1}_{\pm}(\omega,k)$ given by Eq.~(\ref{deltast}) and the real and 
imaginary parts of $\Pi_{\pm}(\omega)$ can be read off from equations
(\ref{imaggauge})-(\ref{imagscalar}).  

The poles of the full fermion propagator determine the excitations 
in the medium. The position of the poles are obtained from
the zeros of $\overline{\Delta}^{-1}_{\pm}(\omega,\vec{k})$ for 
$\omega=\omega_{\text{p}}(k)-i\Gamma(k)$. In the narrow width
approximation $\Gamma(k) \ll \omega_{\text{p}}(k)$, the position of the complex poles are
determined by the following equation
\begin{equation}
{}^{\ast}\Delta_{\pm}^{-1}(\omega_{\text{p}}(k),\vec{k})
+\Pi_{\pm,R}(\omega_{\text{p}}(k),\vec{k}) - i\left[
Z^{-1}_{\pm}(\omega_{\text{p}}(k)) \; \Gamma(k)+
\text{sgn}(\Gamma(k)) \;\Pi_{\pm,I}(\omega_{\text{p}}(k),\vec{k})\right]=0~,
\end{equation}
where we have made use of Eqs.~(\ref{imaggauge}) and (\ref{imagscalar}), and 
$$
Z_{\pm}(\omega_{\text{p}}(k))=\left[\frac{\partial ~ 
{}^{\ast}\Delta_{\pm}^{-1}(\omega,\vec{k})}{\partial \omega}
\right]^{-1}_{\omega=\omega_{\pm}(k)}
$$
are the residues at the poles for the collective excitations.

The real and imaginary parts of the above equation read
\begin{eqnarray}
&&{}^{\ast}\Delta_{\pm}^{-1}(\omega_{\text{p}}(k),\vec{k})
+\Pi_{\pm,R}(\omega_{\text{p}}(k),\vec{k})=0~,\label{realpoleex}\\
&&\Gamma(k)+\text{sgn}\left(\Gamma(k)\right)\; Z_{\pm}(\omega_{\text{p}}(k)) \;
\Pi_{\pm,I}(\omega_{\text{p}}(k),\vec{k})=0~.\label{imagipoleex}
\end{eqnarray}

To leading order in HTL [i.e., neglecting $\Pi_{\pm,R}$ in Eq.~(\ref{realpoleex})]
the real part of the poles are given by the dispersion relations of collective excitations  
\begin{equation}
\omega_{\text{p}}(k)=\left\{
\begin{array}{l}
\pm\omega_{+}(k)\quad\text{ for fermion}~,\\
\pm\omega_{-}(k)\quad\text{ for plasmino}~.
\end{array}\right.\label{realpole2}
\end{equation}
The solution for the imaginary part is obtained by replacing 
$\omega_{\text{p}}(k)$ with (\ref{realpole2}) in
$\Pi_{\pm,I}(\omega_{\text{p}}(k),\vec{k})$ to this order.  

However, the equation for the imaginary part (\ref{imagipoleex}) 
does not have a solution because 
$\Pi_{\pm,I}(\omega_{\text{p}}(k),\vec{k})>0$ (see Eqs.~(\ref{gammaqcd}), 
(\ref{pisfin}) and the discussion below it) 
and $Z_{\pm}(\omega_{\text{p}}(k))>0$. 
Therefore, there is {\em no} complex pole in the physical sheet. This
is a well-known  result: if the imaginary
part of the self-energy on shell  is positive there is no complex
pole solution in the {\em physical} sheet, the pole has moved off into the
unphysical (second or higher) sheet. If the imaginary part on shell 
is negative, there are {\em two} complex poles in the physical sheet corresponding 
to a growing and a decaying solution, i.e., an instability. 
It will become clear below that $\Pi_{\pm,I}(\omega_{\text{p}}(k),\vec{k})>0$ 
corresponding to a
complex pole in an unphysical sheet. In this case the spectral density features a 
Breit-Wigner resonance shape with a width given by the damping rates 
\begin{equation}
\Gamma_{\pm}(k)= Z_{\pm}(k) \;
\Pi_{\pm,I}(\omega_{\pm}(k),\vec{k})~,\label{dampingrates}
\end{equation}
which in general determines an exponential fall-off of the fermion propagator in real time: 
the amplitude for collective excitations with $\chi=\pm 1$ fall off as 
$e^{-\Gamma_{\pm}(k)t}$, an interpretation borne out by the real time
evolution of the initial value problem~\cite{fermion,disip,linon}. An important
exception to this analysis appears in the Abelian gauge theory because
of infrared divergences in the damping rate~\cite{iancu,boyrgir}. 
These will be analyzed in detail below.

\subsubsection{Gauge boson contribution}

We now focus on the contribution to $\Gamma_{\pm}(k)$ from
the soft loop momentum region of the gauge boson exchange. 

For soft momenta $k\approx g\,T$, the non-HTL contribution to the 
fermion self-energy from the gauge boson, i.e., ${}^{\ast}\Sigma_{r\vec k}(\omega)$ 
requires HTL resummed internal propagators and vertices and is subleading by one power of $g$, 
a thorough study is presented in~\cite{brapis,pisarski}. The main ingredients
for a consistent computation of the non-HTL contribution are: (i) the HTL resummed internal 
propagators, (ii) the HTL resummed vertices~\cite{brapis,pisarski,rebhan1,iancu}. 
The resummed propagators are obtained from the spectral
representation in terms of the  HTL spectral densities both for fermions and gauge bosons 
(longitudinal and transverse). 
When all of the lines at the vertex  are soft the differential form of the 
Ward identities from the HTL resummed fermion self-energy can be used to obtain 
the resummed vertex~\cite{brapis,pisarski,rebhan1,iancu}.
We refer the reader to Refs.~\cite{brapis,pisarski,rebhan1,iancu} for the details. 

The analysis leading to the expression of the damping rates
Eq.~(\ref{dampingrates}) relies on the existence and
smallness of the imaginary parts of the self-energy on the mass shell
of the collective excitations. However in gauge
theories potential finite temperature  infrared divergences could
invalidate these conclusions. The infrared divergences
associated with the exchange of longitudinal gauge bosons are a result
of small angle Coulomb scattering and at zero temperature
are those of Rutherford scattering. At finite temperature the
longitudinal gauge boson (instantaneous Coulomb  interaction)
is screened by the Debye screening mass $m_D \propto g\;T$ which cuts
off the infrared and  leads to a finite contribution
to the damping rate from longitudinal gauge bosons. For soft external
momentum this contribution has been found to be given by~\cite{pisarski,iancu} 
\begin{equation}
\Gamma^{l,g}(k)= \alpha\;  A\;T~,\label{longi}
\end{equation}
with $A$ a constant that can be found numerically~\cite{pisarski} and in QCD is 
of $ {\cal O}(1)$ and $\alpha = g^2/4\pi$. 

For a fermion excitation at {\em rest} (fermions and plasminos
coincide for $k=0$),  
the damping rate has been computed~\cite{brapis,pisarski,rebhan1} and
found to be given by 
\begin{equation}
\Gamma^{g}(k=0) = \alpha\; C\;T~,  \label{restgamma}
\end{equation}
with $C$ again being a numerical constant of ${\cal
O}(1)$~\cite{brapis,pisarski}. The reason 
that we do not specify the constants quantitatively is because we are
interested in the damping rate for {\em moving} collective excitations. When
the collective excitations are at rest there is no difference in their
dispersion relations and therefore their transport properties are identical.

In QCD (and also in the SU(2) sector of the Standard Model) the
potential infrared divergence arising from  the exchanged transverse
gluon propagator is conjectured to be screened by  
the non-perturbative magnetic screening mass $m_{mag} \approx g^2 T$ 
($g$ is the QCD coupling constant) and the longitudinal gluon is Debye
screened with $m_D \approx g\,T$.  
Thus the infrared divergences are cured by electric and magnetic
screening lengths. 
The detailed analysis of Refs.~\cite{brapis,pisarski,rebhan1} 
lead to the result that the damping rate of moving quasiparticles 
with momenta $k\gg  g^2 T$ in a non-abelian plasma are given by 
(up to an overall constant that depends on the gauge group structure)
\begin{equation}
\Gamma^g_{\pm}(k) = \frac{g^2T}{4\pi}|v_{\pm}(k)|\;
\ln\frac{1}{g}~,\label{gammaqcd} 
\end{equation}
where $v_{\pm}(k)$ are the group velocities for the fermion and plasmino
branches. This expression is {\em not} valid for $k=0$ where the damping
rates of fermion and plasmino at rest do not have the logarithmic behavior 
in terms of the gauge coupling~\cite{brapis,pisarski},
nor near $k=k_{min}$ where the group velocity of the plasmino branch 
vanishes~\cite{conversation}. 


We point out that our results for the mean free paths of the collective excitations and
their {\em difference} will be restricted to the domain of validity of Eq.~(\ref{gammaqcd}), 
furthermore we emphasize that this regime of validity has been pointed out in 
Refs.~\cite{pisarski,rebhan1,iancu,conversation}). 
In the region where the plasmino group velocity vanishes there are  
contributions to its damping rate that must 
be understood in detail, such a task is  clearly beyond
the scope of this article. 

Since the HTL structure in QED is similar 
(up to gauge group factors) to that of QCD (and scalar QED), 
Pisarski~\cite{pisarski} suggested that a similar form of the damping rate 
should be valid in a QED plasma, despite
the fact that there is no magnetic screening in the Abelian theory. 

In QED the transverse photon propagator is only {\em dynamically} screened 
via Landau damping and the infrared divergences remain, possibly to all
orders in perturbation theory. These divergences led to questioning the
quasiparticle interpretation of moving charged excitations~\cite{kobes,pilon}. More
recently~\cite{iancu} a detailed study of the fermion propagator in the
Bloch-Nordsieck 
(eikonal) approximation in {\em real time} revealed that for $\omega_D\;t\;
|v_{\pm}(k)| \gg 1$   
\begin{equation}
S_k(t) \approx \exp[-\alpha T |v_{\pm}(k)|\; t \ln(\omega_D\;t\;
|v_{\pm}(k)|)]~,  \label{nonexpo} 
\end{equation}
with $\omega_D \approx g\,T$ the Debye frequency and $v_{\pm}(k)$ the
group velocity  
of the fermion and plasmino branches. Although this
is not an exponential relaxation that would emerge from a Breit-Wigner
resonance shape of the spectral density as argued above, it does reveal
a particular time scale from which a damping rate  can be extracted
and is given by 
\begin{equation}
\Gamma^g_{\pm}(k) \approx \alpha\; T\; |v_{\pm}(k)|
\ln\frac{1}{g}~.\label{infrarate} 
\end{equation} 

This result has also been recently found in scalar QED with an
alternative method  based on the renormalization group 
and within the context of an initial value problem as followed
here~\cite{boyrgir}.  Since SQED, QED and QCD share
the same HTL structure to lowest order~\cite{rebhan,lebellac}, the
analysis via the renormalization group furnishes an independent 
confirmation of the eikonal approach. 

At this stage it is important to describe the physics that leads to
the damping rate from the gauge  
boson contribution, which as it will be seen below is rather different
from that of the scalar.  
The constraints
from energy momentum conservation in the imaginary part of the
self-energy can only be satisfied 
below the light cone where the HTL resummed gauge boson propagator has
support that arises from  
the Landau damping cut. 
The infrared process that leads to the non-exponential relaxation is
the emission  and absorption of soft photons or gluons at almost right
angles with the moving fermion~\cite{pisarski}.  

Thus we summarize the contribution to the damping rate of the
collective excitations by the HTL resummed gauge boson exchange:
\begin{eqnarray}
&& \Gamma^g_{\pm}(k) \approx \alpha\; T\; |v_{\pm}(k)|\;
\ln\frac{1}{g}\quad \text{for} 
\quad k \gg g^2T~,\label{movingrate} \\
&& \Gamma^g_{\pm}(k)\approx\alpha\; T\quad\quad\quad\quad\quad\quad~\;
\text{for}\quad k=0~. \label{restrate}
\end{eqnarray}

The damping rate for moving fermions is not yet available for the whole range of group 
velocities and clearly a better understanding of the infrared region must be pursued. In particular
Eq.~(\ref{movingrate}) is not valid near the plasmino minimum when
the group velocity vanishes, since subleading contributions become important
in this region~\cite{conversation}. 
Fig.~2 displays $\Gamma^g_{\pm}(k)/M_{eff}$ vs $k/M_{eff}$ for $g=0.3$.

\subsubsection{Scalar contribution}

In Ref.~\cite{fermion}
the damping rate for a fermion in a scalar plasma was derived. There it
was found that the {\em decay} of the scalar into fermion-antifermion pairs results 
in a quasiparticle width for the fermion. 

Here we study a different aspect, that is the lifetime of the {\em collective excitations} 
both fermions and plasminos due to the process of
scalar decay into the collective excitations. This is one of the novel contributions of this article. 

Writing the Laplace transform of the scalar contribution to the
self-energy as in  Eq.~(\ref{epsilons2}), the
scalar contribution to $\Pi_{\pm}(\omega,k)$ in Eq.~(\ref{totalpi})
are given by [see Eq.~(\ref{epsilons2}) for the definition of the
$\varepsilon^{(i)}_{\vec k}(\omega)$] 
\begin{equation}
\Pi^s_{\pm}(\omega,k) = \omega \;  \varepsilon^{(0)}_{\vec k}(\omega) \pm k \;
\varepsilon^{(1)}_{\vec k}(\omega)~.
 \label{piscalar}
\end{equation}

The full expressions for the $\varepsilon^{(i)}_{\vec{k}}(\omega)$ to one loop 
are given in Ref.~\cite{fermion}, we only quote here  the most relevant 
features of their imaginary parts to clarify our arguments. 
The imaginary parts of these coefficients are given by
\begin{eqnarray}
\varepsilon^{(0)}_{I,\vec{k}}(\omega) & = &
\frac{\pi}{2|\omega|}\mbox{sgn}(\omega)\left[ 
\rho^{(0)}_{\vec{k}}(|\omega|)+\rho^{(0)}_{\vec{k}}(-|\omega|)\right]~,
\nonumber \\ 
\varepsilon^{(1)}_{I,\vec{k}}(\omega) & = &
\frac{\pi}{2}\mbox{sgn}(\omega)\left[ 
\rho^{(1)}_{\vec{k}}(|\omega|)-
\rho^{(1)}_{\vec{k}}(-|\omega|)\right]~,
\label{imagipartepsi}  
\end{eqnarray}
in terms of the following one-loop spectral densities~\cite{fermion}
\begin{eqnarray}
\rho^{(0)}_{\vec{k}}(\omega)&=& y^2 \int \frac{d^3 q}{(2\pi)^3
}\frac{\bar{\omega}_{q}}{
2 \omega_{k+q} \bar{\omega}_{q}}\nonumber\\
&& \times\left[\delta(\omega-\omega_{k+q}-\bar{\omega}_{q})
(1+n_{k+q}-\bar{n}_{q})+ \delta(\omega-\omega_{k+q}+\bar{\omega}_{q})
(n_{k+q}+\bar{n}_{q}) \right]~,
\nonumber\\ \rho^{(1)}_{\vec{k}}(\omega)&=& y^2 \int \frac{d^3 q}{(2\pi)^3
} \frac{1}{2 \omega_{k+q} \bar{\omega}_{q}}\frac{\vec k \cdot \vec
q}{k^2}\nonumber\\ 
&& \times\left[\delta(\omega-\omega_{k+q}-\bar{\omega}_{q})
(1+n_{k+q}-\bar{n}_{q})-\delta(\omega-\omega_{k+q}+\bar{\omega}_{q})
(n_{k+q}+\bar{n}_{q})\right]~,\label{rhos}
\end{eqnarray}
where $\omega_{k+q}=\sqrt{m^2+(\vec{k}+\vec{q})^2}$, and
$n_{k+q}$ and $\bar{n}_q$ are the respective distribution functions for the scalar and the fermion  
in the loop.
The terms that contain the $\delta(\omega-\omega_{k+q}+\bar{\omega}_{q})$ do not give a 
contribution to the damping rate. These arise from the processes $\psi\rightarrow\phi +\psi$ 
(with $\psi$ and $\phi$ denoting the fermion and the scalar, respectively) 
and result in the usual two particle cuts with support 
for $|\omega| > \sqrt{k^2+(m+M)^2}$ (here,
$M$ and $m$ are the respective masses of the fermion and the scalar in the loop). 
Thus we only consider the terms 
proportional to $n_{k+q}+\bar{n}_q$ in 
$\Pi^s_{\pm,I}(\omega,\vec{k})$ since only the delta functions that multiply 
these terms will have support on the
mass shell of the collective excitations as explained below.  
These terms lead to the following imaginary part of the scalar contribution
\begin{eqnarray}
\Pi^s_{\pm,I}(\omega,\vec{k})&=&\pi y^2\int \frac{d^3 q}{(2\pi)^3}
\frac{n_{k+q}+\bar{n}_q}{2\,\bar{\omega}_q\, 2\, \omega_{k+q}} \left[\left(
\bar{\omega}_q \mp \text{sgn}(\omega)\hat{k}\cdot \vec{q}\right)
\delta(|\omega|-\omega_{k+q}+\bar{\omega}_q)\right.\nonumber\\
&&\quad\quad\quad +\left.\left(\bar{\omega}_q \pm \text{sgn}(\omega)\hat{k}\cdot \vec{q}\right)
\delta(|\omega|+\omega_{k+q}-\bar{\omega}_q)\right]~. \label{pisfin}
\end{eqnarray}
It is straightforward to show that $\Pi^s_{\pm,I}(\omega,\vec{k})>0$ from 
the energy conservation constraints: 
$|\omega|\pm\bar{\omega}_q\mp\omega_{k+q}=0$, and therefore this imaginary part is 
not associated with an instability
but with a true resonance, i.e., a complex pole in an unphysical sheet and its real time 
interpretation corresponds
to an exponential fall off of the amplitude of the collective excitation.

The first delta function determines a cut in the region 
$0<|\omega|< \sqrt{k^2+(m-M)^2}$ and originates in the physical process 
$\phi \rightarrow \psi + \bar{\psi}$ whereas the
second delta function determines a cut in the region $0<|\omega|<k$ 
and originates in the process
$\phi + \psi \rightarrow \psi$. 
The first cut describes the  decay of the scalar
into fermion-antifermion pairs,  
the second cut for ( $\omega^2<k^2$) is associated with  Landau damping. 
Both delta functions restrict the range of the integration
variable $q$ (see below). However, since the dispersion relations for
the collective excitations are always above the light cone, i.e., $\omega_{\pm}(k) >k $, 
the contribution from the second delta function vanishes on the mass
shell of the collective excitations. Therefore only the first term  in Eq.~(\ref{pisfin}) 
contributes to the lifetime of the collective excitations.

It is at this stage that we are in position to formalize the arguments 
presented in the previous section 
to justify a one-loop computation of the scalar contribution to the self-energy of the
collective excitations in terms of {\em free field} propagators (non HTL resummed). 

The argument hinges upon two important features of the spectrum of the fermionic collective
excitations: (i) The dispersion relation is always above the light cone, this
feature guarantees that the second cut in (\ref{pisfin}) with support below the light cone will
not contribute to the imaginary part evaluated on the mass-shell of the collective excitation. (ii) 
Under the assumption that the scalar mass $m \gg g\,T$ and an external 
momentum $k \approx g\,T$ and therefore 
$|\omega| \approx g\,T$ on the mass shell of
the collective excitations, the first delta function is satisfied only for 
$\bar{\omega}_q \approx \sqrt{q^2+m^2} \gg  g\,T$,
hence the fermion in the loop is the ordinary single particle fermionic excitation, 
{\em not} the collective excitations associated
with soft momenta. This is clear from the properties of  dispersion relations and 
residues for the collective excitations
displayed in Eq.~(\ref{properties}) since for large momenta only the fermion branch 
survives (the wave function renormalization of the plasmino branch vanishes) and approaches 
the vacuum fermion dispersion relation.
Hence in summary: the contribution from scalar exchange to damping rates of 
the soft collective excitations is obtained from the one loop
self-energy in which the internal scalar {\em and fermion} lines are the free 
particle propagators, but the external
line are the fermion or plasmino collective excitations.   
 
This is a remarkable result, 
{\em the fermions and plasminos acquire a width through the
induced decay of the scalar in the medium} and the damping rates for the soft collective 
excitations are {\em different}. For large scalar mass $m \gg g\,T$ the kinematics dictates 
that the scalar decays into a soft collective excitation, either fermion or plasmino, 
and a hard fermion.

In the limit of $ M\ll g\; T $,  
we finally find 
\begin{eqnarray}
&&\Pi^s_{\pm,I}(\omega,\vec{k})=\frac{y^2}{32 \pi k^2} 
\int^{q^{\ast}_2(\omega)}_{q^{\ast}_1(\omega)} dq\; 
(n_{k+q}+\bar{n}_q) \left[ 2kq \mp \text{sgn}(\omega) \;
(2\; |\omega|\; q + \omega^2-\omega^2_{k})\right] \label{finparteimaginaria}
\end{eqnarray}
for $|\omega|>k$, where $\omega_k=\sqrt{m^2+\vec{k}^2}$ and
$$q^{\ast}_1(\omega)=\frac{1}{2}\left|\frac{\omega^2-\omega^2_k}{|\omega|+k}\right|~,\quad
q^{\ast}_2(\omega)=\frac{1}{2}\left|\frac{\omega^2-\omega^2_k}{|\omega|-k}\right|~.
$$

From the expression for the total damping rates Eq.~(\ref{dampingrates}),
the scalar contribution to the damping rates for the fermion and
plasmino collective excitations is found by evaluating the scalar
contribution of the imaginary part of the  
self-energy (\ref{pisfin})-(\ref{finparteimaginaria}) on the mass
shell of the collective modes. We find the damping rates to be given
by  
\begin{eqnarray}
\Gamma^s_{\pm}(k)&=&Z_{\pm}(k)\;\Pi^s_{\pm,I}(\omega_{\pm}(k),\vec{k})\nonumber\\
&=&\frac{y^2 Z_{\pm}(k)}{32\pi k^2}\Biggl\{\pm
T \left( \omega_k^2 - {\omega^2_{\pm}(k)} \right) 
\left[\ln(1 - e^{-\beta[\omega_{\pm}(k)+q]}) - \ln(1 + e^{-\beta q})\right]\nonumber\\
&&\pm 2 T\left(\omega_{\pm}(k) \mp k\right) \left[       
\omega_{\pm}(k) \ln (1 - e^{-\beta[\omega_{\pm}(k)+q]}) + 
q \ln (1 + e^{-\beta q}) \right] \nonumber\\
&& \mp \left.2 T^2 \left(\omega_{\pm}(k) \mp k \right) \left[ 
\text{Li}_2(1 - e^{-\beta[\omega_{\pm}(k)+q]}) + 
\text{Li}_2(-e^{-\beta q}) \right]\Biggr\}
\right|^{q^{\ast}_{2}(\omega_{\pm}(k))}_{q^{\ast}_{1}(\omega_{\pm}(k))}~,
\label{scalarrate} 
\end{eqnarray}
respectively. Here, 
$$\text{Li}_2(x)\equiv\int^{0}_x \frac{dt}{t}\ln(1-t)$$
is the dilogarithm function.

Although these expressions are somewhat unwieldy, there are important features that are very 
revealing in Eq.~(\ref{scalarrate}): these
are the wave function renormalization factors $Z_{\pm}(k)$. 
As we discussed in the previous section and is clear from the expressions 
(\ref{properties}) the wave function renormalization for the plasmino
branch vanishes very fast for $k \geq k_{min} \approx 0.4\; M_{eff}$. 
The damping rates $\Gamma^s_{\pm}(k)$ are 
displayed in Fig.~3 vs $k/M_{eff}$ for $g=0.3$ and $m=T$. The upper solid
line is the damping rate for the fermion branch, and the lower dashed line is the damping 
rate for the plasmino branch, 
the rapid decrease for the plasmino branch beyond $k \geq M_{eff}$ is a result of the
vanishing of the wave function renormalization. 

\section{Mean free paths: a wave packet interpretation}

The notion of mean free paths in a plasma in which different collective
excitations with different group velocities are present is rather subtle and requires a careful 
treatment. It is here where the real time description in terms of the in-medium 
Dirac equation is fruitful. We  obtain the mean 
free paths by studying the space-time evolution of an initially prepared wave packet. 
This formulation identifies unambiguously the mean free
paths without resorting to a diffusion interpretation of transport.

In particular we are interested in transport of chirality by the collective excitations, 
which evolve independently in time. 
Since the collective excitations  are eigenstates of the effective Dirac equation in the 
leading HTL order, they are given by the solutions of 

\begin{equation}
\left[{}^{\ast}\Delta^{-1}_{-}(\omega,\vec{k})(\gamma_0-
\vec{\gamma}\cdot\hat{k})+{}^{\ast}\Delta^{-1}_{+}(\omega,\vec{k})
(\gamma_0+\vec{\gamma}\cdot\hat{k})\right]\tilde{\psi}_{\vec{k}}(s=-i\omega)=0~.
\end{equation}

Let us consider
that the initial expectation value $\psi_{\vec{k}}(0)$ in Eq.~(\ref{laplakern}) is of the form
\begin{equation}
\psi^{\pm}_{\vec{k}}(0)\; \propto \; {\cal U}^{\pm}_{\vec k}~
 e^{-\frac{R^2}{2}(\vec k-\vec{k}_o)^2}~, \label{inipacket}
\end{equation}
where the spinors ${\cal U}^{\pm}_{\vec k}$ are annihilated by 
$\gamma_0\mp \vec \gamma 
\cdot \hat k$, i.e., 
have chirality to helicity ratios $\chi=\pm 1$ respectively. 
This initial state describes a wave packet of collective modes localized in momentum 
at $\vec{k}_o$. In space it describes a wave packet localized
over a spatial extent $R$ around the origin
with expectation value of the momentum $\vec{k}_o$. To describe
an almost monochromatic beam of collective excitations we will choose $R\gg \omega^{-1}_{\text{p}}$ 
since $\omega^{-1}_{\text{p}}$ is the typical spatial extent
of the quasiparticle states (screening cloud). The quasiparticle is
described by a spectral function of the Breit-Wigner form (although this
is not evident for the logarithmic relaxation in QED, it has been argued
to be a good approximation in this case~\cite{iancu}). The time evolution
can then be obtained by inverse Laplace transform, the details of which
can be found in Ref.~\cite{linon}. In the narrow width approximation 
$\Gamma^t_{\pm}(k) \ll  \omega_{\pm}(k)$,  we find the following time evolution of the 
initial wave packet 
\begin{equation}
\psi^{\pm}(\vec x,t) \approx \int d^3 k \; e^{i\vec k \cdot \vec x} \;
 Z_{\pm}(k) \; {\cal U}^{\pm}_k ~ \;
 e^{-\frac{R^2}{2}(\vec k-\vec{k}_o)^2} \; e^{-i\omega_{\pm}(k)t} \;
 e^{-\Gamma^t_{\pm}(k) \;t}~, \label{timedeppack}
\end{equation}
where $\Gamma^t_{\pm}(k)$ is the {\em total} damping rate, i.e., 
gauge plus scalar contribution and $Z_{\pm}(k)$ the residue at the quasiparticle pole 
(wave function renormalization). 
Since the initial wave packet is
strongly peaked in momentum, we can perform the integral over momentum by
expanding $\omega_{\pm}(k)$ and $\Gamma^t_{\pm}(k)$ around 
$k=k_o$ and using the narrow width approximation. We find
\begin{equation}
\psi^{\pm}(\vec x,t) \propto Z_{\pm}(k_o)\; {\cal U}^{\pm}_{\vec{k}_o}~~ 
e^{-\frac{\vec{X}^2_{\pm}(t)}{R^2(t)}}~ e^{-\Gamma^t_{\pm}(k_o) t}~ 
e^{i\vec{k}_o \cdot(\vec x - \vec{V}_{p,\pm}~ t)}~,
\end{equation}
where 
\begin{eqnarray} 
\vec{X}_{\pm} (t) & = &  \vec x - \vec{V}_{g,\pm}\; t~,\quad 
R^2(t) =  R^2 + 
\left. i~\frac{d^2 \omega_{\pm}(k)}{dk^2}\right|_{k_o}\; t~,\\
\vec{V}_{g,\pm} & = &\left. \hat{k}_o~\frac{d \omega_{\pm}(k)}{dk}\right|_{k_o}  
= \hat{k}_o v_{\pm}(k_o)~, \label{groupvelo}\\
\vec{V}_{p,\pm} & = & \hat{k}_o \frac{\omega_{\pm}(k_o)}{k_o} \label{phasevelo}~.
\end{eqnarray}
Obviously $\vec{V}_{g,\pm}$ and $\vec{V}_{p,\pm}$ are the group and
phase velocities, respectively. We then see that the peak amplitude of the
wave packet attenuates in space on a distance scale given by the {\em
mean free paths} 
\begin{equation}
\lambda_{\pm}(k_o) = \frac{|v_{\pm}(k_o)|}{\Gamma^t_{\pm}(k_o)}~,
\label{meanfreepaths} 
\end{equation}
\noindent where $v_{\pm}(k)$ are the group velocities for the fermion and
plasmino branches and we took the absolute value because there is a 
region of negative group velocity for the plasmino branch (see Fig.~1) before
the minimum at $k \approx 0.4\; M_{eff}$. Eq.~(\ref{meanfreepaths}) is the
main reason for the wave packet analysis: it unambiguously identifies
the mean free paths of the collective excitations in terms of the
group velocities and  
damping rates for each branch of collective excitation independently. 

Since the gauge contribution to the damping rates of collective excitations
is proportional to their group velocities (see Eq.~(\ref{movingrate})) we then
highlight the fact that {\em whereas the gauge contribution to the damping
rates of the collective modes are different, their mean free paths are
the same}. Hence the {\em difference of mean free paths} which is the
goal of our study is solely given (to this order) by the scalar contribution.
This is the main result of this study. 

Using the expression for the gauge contribution to the damping rate of moving 
quasiparticles given by Eq.~(\ref{movingrate}), we can
write the mean free paths in the following form
\begin{equation}
\lambda_{\pm}(k) = \frac{1}{\Gamma^g_{o}}
\left[1+\frac{\Gamma^s_{\pm}(k)}{|v_{\pm}(k)|\;
\Gamma^g_{o}}\right]^{-1}~,\quad 
\Gamma^g_o \approx \alpha \; T\;  \ln\frac{1}{g}~. \label{meanfre}
\end{equation}

This expression reveals clearly that the {\em difference}
of mean free paths for the fermion and plasmino collective excitations is
mostly given by the contribution to the damping rate from the {\em
scalar sector}.  
This is one of the important results in this article that we want
to emphasize: although the largest contribution to the damping rates for soft 
collective excitations arises from the gauge boson contribution to the
self-energy,  
since $\alpha \gg  y^2$, the {\em difference} of mean free paths is determined 
mainly by the decay of the heavy scalar.  
Fig.~4 shows $[\lambda_{+}(k) - \lambda_{-}(k)]M_{eff}$ vs $k/M_{eff}$
for $g=0.3$,  
$y=10^{-4}$ and $m=T$. This figure displays one of the important
results of this article.  
The strong peak near the plasmino minimum is a result of the vanishing
of the plasmino  
group velocity and as discussed previously the gauge contribution to
the damping rate of  
the plasmino branch is not reliable near this region but is so beyond and
before this point. 

In particular away from the minimum from the plasmino branch and using
$y^2\ll \alpha$, the difference in mean free paths is approximately given by
\begin{equation}
\lambda_+(k)-\lambda_-(k) \approx
\frac{1}{(\Gamma^g_o)^2}\left[\frac{\Gamma^s_-(k)}{|v_-(k)|}-  
\frac{\Gamma^s_+(k)}{v_+(k)}\right]~. \label{differencemean} 
\end{equation}

Thus the fermion and plasmino soft collective excitations have
different mean free paths and the  
difference is determined mostly by the heavy scalar decay 
into a hard and a soft collective excitation. 

\section{Summary, implications for the Standard Model and conclusions}

We have focused our attention on the transport properties of soft collective excitations 
in an abelian gauge, fermion and scalar plasma with the goal of
understanding chirality transport as relevant for non-local baryogenesis. 

Our main observation is that the transport properties of {\em soft} collective excitations 
are different for the different branches of the fermionic
dispersion relation. Since these collective excitations carry ratios of chirality
to helicity $\chi=\pm 1$, different transport coefficients, i.e., damping rates
and mean free paths for these excitations will result in a differential
transport of chirality. There are two main physical mechanisms that lead
to different transport properties: (i) The absorption and emission of soft
gauge bosons that are dynamically screened by Landau damping on scales $g\,T$ and by 
a magnetic mass for the non-abelian sector on scales $g^2T$. (ii) The 
{\em decay} of the heavy scalar (Higgs) particle into hard fermion and a soft fermion 
or plasmino results in a contribution to the damping rate of on-shell collective excitations. 
The physical aspects of these two phenomena are fairly robust, in particular the 
hard thermal loop structure of the fermionic self-energy has
the {\em same form} in the abelian (QED) and non-abelian (SM) theories,
with the only differences being in the overall scales (see for example Ref.~\cite{farrar} 
wherein the contributions of the weak and scalar sectors to the leading HTL fermion self-energy 
had been computed in detail). The decay of a heavy
Higgs into a  hard fermion and a soft collective excitation is a consequence
of simple kinematics in the heat bath and is therefore fairly independent
of the model. Obviously the strength of the Yukawa couplings and the group
representation of the scalar fields will change the overall normalization
of the scalar contribution. 

Therefore we believe that the results obtained in this article are relevant for non-local 
baryogenesis in the Standard Model or extensions thereof and that the main ingredients 
that determine the different transport coefficients described in this
study are fairly robust. 

We have combined detailed studies of the damping rates of moving fermionic 
quasiparticles~\cite{brapis,pisarski,rebhan1,iancu} and extended the recently 
studied phenomenon of fermionic damping rates via heavy scalar decay~\cite{fermion} 
to provide a description 
of the damping rates and mean free paths of collective excitations with momenta 
$k \gg g^2 T$ to leading order in Hard Thermal Loop 
resummation and lowest order in Yukawa coupling. This
analysis is valid for the lightest quarks and leptons in the Standard Model
(certainly {\em not} the top quark) and both in the symmetric and broken symmetry
phases, since for the lightest fermions $g\,T \gg M$ with $M$ the vacuum masses. 
The main uncertainties in our results stem from the infrared sensitivity of 
the gauge contributions to the fermionic
damping rate, in particular the magnetic sector and its non-perturbative
screening scale. This is still an ongoing subject of active 
study~\cite{pisarski,rebhan1,iancu} and certainly beyond the scope of this
article. However as argued  persuasively in~\cite{pisarski,rebhan1,iancu}
the results given by Eqs.~(\ref{infrarate}) are trustworthy for $k\gg g^2T$ 
and away from
the minimum in the plasmino branch when the plasmino group velocity vanishes. 
For a first order phase transition the Higgs mass does not vary
much near the phase transition from its vacuum value 
$m \approx T \approx 100 \mbox{GeV}$. This translates into that the scalar 
contribution to the fermionic self-energy is not sensitive to the infrared 
behavior and does not require hard-thermal loop resummation. 

By casting our studies in terms of the real time, in-medium effective 
Dirac equation, we were able to study the real 
time evolution of wave packets of
collective excitations and extract unambiguously the mean free paths or attenuation 
lengths for the different collective excitations. This approach transcends any approximation 
that relies on a diffusive description of transport. 

Our main results are summarized by  the expressions: (\ref{scalarrate})
for the scalar contribution to the damping rate  and (\ref{meanfre}) 
for the mean free paths for the soft collective excitations
with $k\gg  g^2 T$ and away from the plasmino minimum $k \approx 0.4\; M_{eff}$. 
The main observation is that the  {\em difference}
of mean free paths for the fermion and plasmino branches for $k\gg g^2T$ and
away from the minimum of the plasmino branch is 
approximately given by Eq.~(\ref{differencemean}), i.e., {\em mainly} determined by 
the different {\em scalar} contributions to the  damping rate of the collective excitations. 
The uncertainties near the region of the
plasmino minimum are related to the infrared sensitivity of the fermionic
damping rates and require a careful analysis of the infrared region of the
HTL resummed self-energies, such a study is currently under way. 

There is potentially an important exception to the validity and robustness of our 
results in the case of 
extensions of the Standard Model that include scalars that are too light. 
Light scalars are common in supersymmetric extensions and extensions with more than one doublet. 
They are welcome because they tend to increase the strength of
the first order phase transition~\cite{trodden}. If light scalars are present, 
their contribution to the fermionic
self-energy will have to be understood in detail and assess in each particular 
model whether an HTL resummation of
the scalar propagator as well as that of the fermion in the loop is necessary. 
Clearly such cases must be analyzed in detail for the particular models. 

A very important limitation of our results is the knowledge of the damping rate of the
plasmino near the region where its group velocity vanishes. As recognized by the 
authors~\cite{pisarski,rebhan1,iancu,conversation}
of the original work, in this region there are further contributions to the 
plasmino damping rate that must be
understood in detail.

We think that the novel results found in this article for the transport of chirality by
soft collective excitations reveal
new features which are distinct from those previously obtained for hard excitations 
 and could have an important impact in mechanisms of local baryogenesis. Whereas the main
focus of this article has been a quantitative understanding of chirality transport by {\em soft} 
collective excitations and a detailed understanding of the gauge and scalar contributions 
to the transport coefficients, obviously
the next step is to provide a quantitative assessment of  the impact of these 
new results on the baryon asymmetry within particular extensions of the Standard Model. 
This study is currently underway.

\section{Acknowledgments}
D.B.~thanks Robert Pisarski and Edmond Iancu for fruitful and illuminating 
comments and explanations. D.B.~and S.-Y.W.~thank the NSF for partial
support through grants PHY-9605186, INT-9815064 and INT-9905954. 
S.-Y.W.~acknowledges support through the Andrew Mellon Predoctoral Fellowship.  
D.B.~and H.J.d.V.~acknowledge support from NATO. 
D.-S.L.~thanks support from the Republic of China National Science Council 
through grant NSC88-2112-M-259-001. 
Y.J.N.~thanks DOE for partial support through grant DE-FG05-85ER-40219 Task A.




\newpage


\begin{center}
\begin{figure}[t] 
\epsfig{file=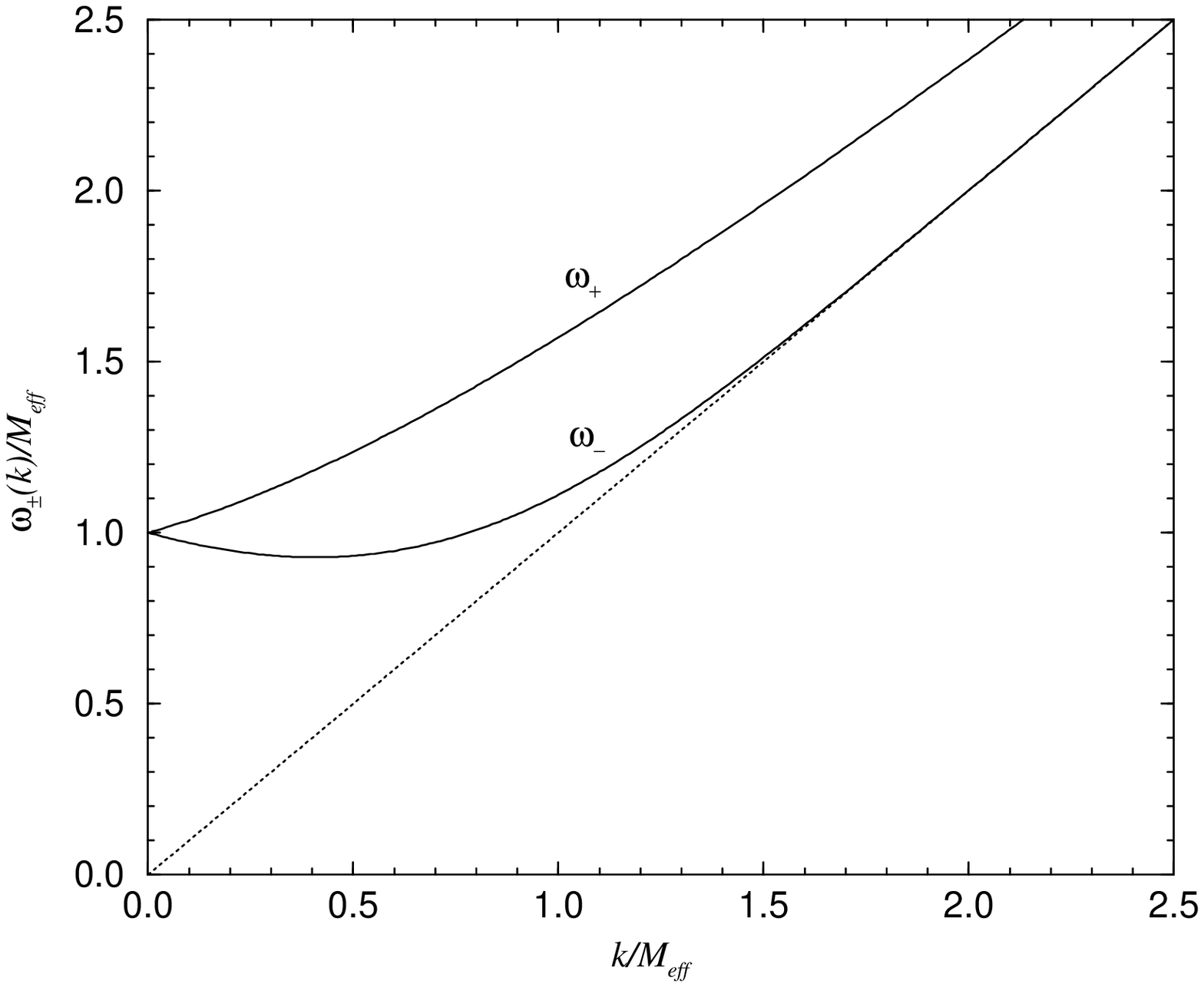,width=10cm,height=10cm} 
\caption{Dispersion relations: the upper correponds to the fermion branch and 
the lower to the plasmino
branch. The dashed line is the light cone.  \label{fig1}}
\end{figure} 
\end{center}

\begin{center}
\begin{figure}[t] 
\epsfig{file=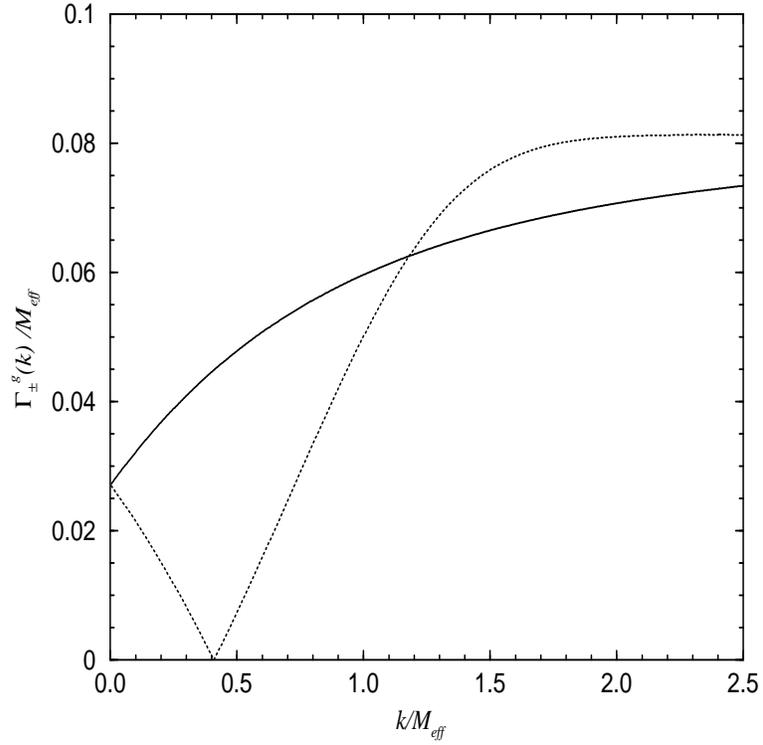,width=10cm,height=10cm} 
\caption{$\Gamma^g_{\pm}/M_{eff}$ vs $k/M_{eff}$ for $g=0.3$.
The solid line corresponds to the fermion
branch and the dashed line to the plasmino branch.  \label{fig2}}
\end{figure} 
\end{center}

\begin{center}
\begin{figure}[t]
\epsfig{file=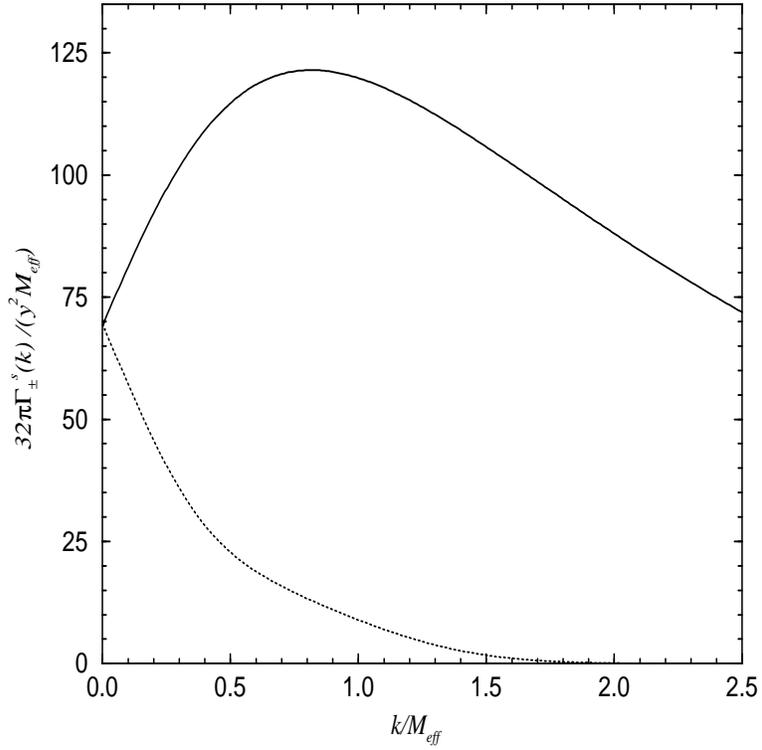,width=10cm,height=10cm}
\caption{$32\pi\Gamma^s_{\pm}(k)/(y^2 M_{eff})$ vs $k/M_{eff}$ for $m=T$. 
The solid line is for the fermion
branch and the dashed line for the plasmino branch. 
The scalar contribution $\Gamma^s_{\pm}(k)$ is given by
Eq.~(\ref{scalarrate}) in the text.\label{fig3}}
\end{figure}
\end{center}

\begin{center}
\begin{figure}[t]
\epsfig{file=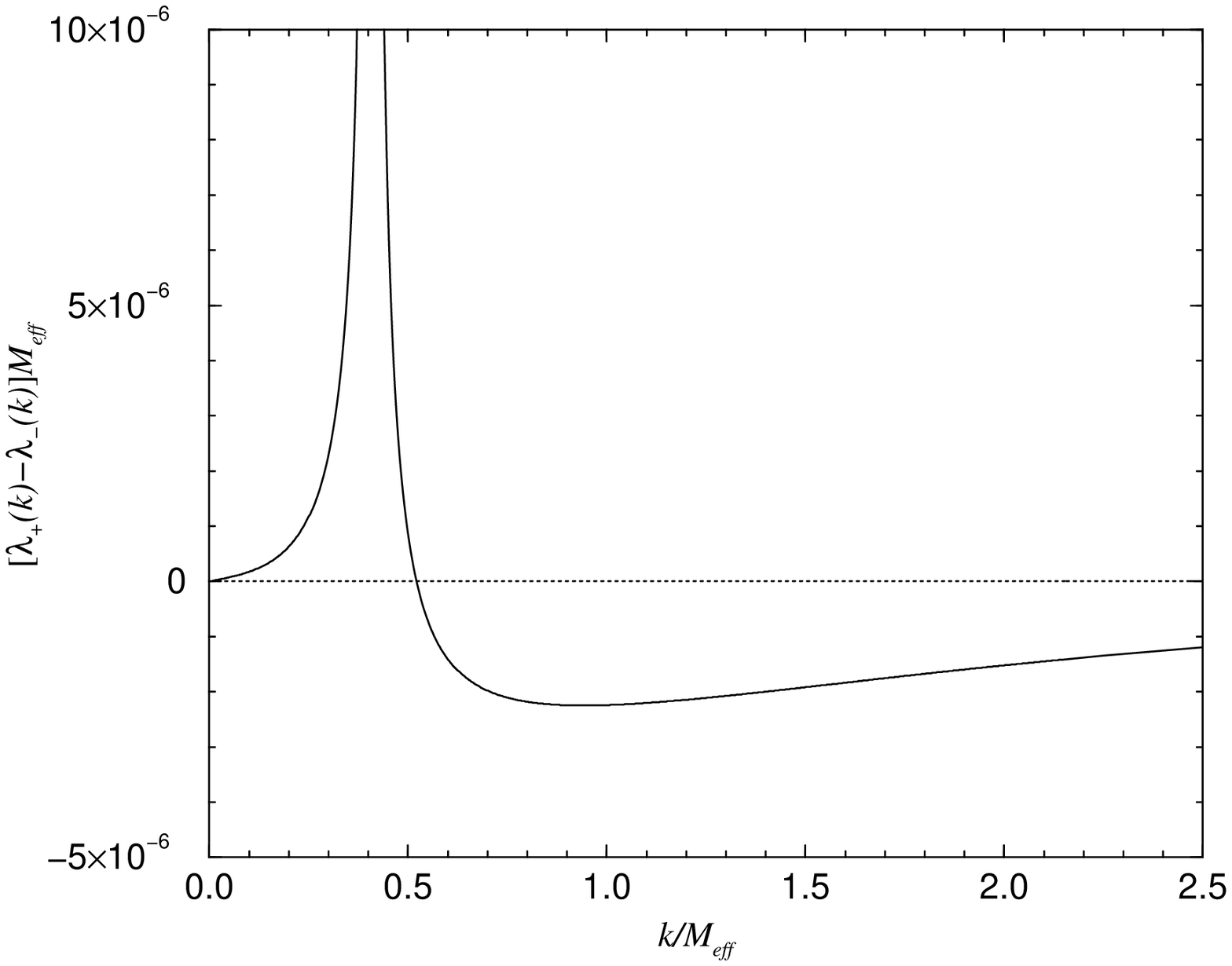,width=10cm,height=10cm}
\caption{ $[\lambda_{+}(k) - \lambda_{-}(k)]M_{eff}$ vs $k/M_{eff}$ for $g=0.3$, 
$y=10^{-4}$ and $m=T$. The peak near $k\approx 0.4~M_{eff}$ 
is a consequence of the vanishing of the plasmino group 
velocity, the region near this peak is not trustworthy as 
the expression for the gauge contribution to
the damping rate of the plasmino is not valid in the region where 
its group velocity vanishes as explained
in the text below Eq.~(\ref{gammaqcd}) . \label{fig4} }
\end{figure}
\end{center}


\begin{thebibliography}{99}
\bibitem{kuzmin} V. A. Kuzmin, V. A. Rubakov and M. E. Shaposhnikov, Phys. Lett. 
B {\bf 155}, 36 (1985). 
\bibitem{dolgov} For a review of possible mechanisms for Baryogenesis, see
A. Dolgov, Phys. Rep. {\bf 222}, 309 (1992). 

\bibitem{turok} N. Turok, in {\em Perspectives on Higgs Physics}, 
edited by G. L. Kane (World Scientific, Singapore, 1993). 

\bibitem{cohen} A. G. Cohen, D. B. Kaplan and A. E. Nelson, Ann. Rev. Nucl. Part. Sci. {\bf 43}, 
27 (1993); Phys. Lett. {\bf B263}, 86 (1991); 
Phys. Lett. {\bf B295}, 57 (1992); Nucl. Phys. {\bf B349}, 727 (1991). 

\bibitem{ruba} V. A. Rubakov and M. E. Shaposhnikov, Phys. Usp. {\bf 39}, 461 (1996). 

\bibitem{aleph} ALEPH Collaboration, Phys. Lett. {\bf B440}, 403 (1998). 

\bibitem{csikor}  F. Csikor, Z. Fodor, J. Heitger, Y. Aoki, A. Ukawa, hep-ph/9901307.  

\bibitem{farrar} G. R. Farrar and M. E. Shaposhnikov, Phys. Rev. D {\bf 50}, 774 (1994); 
G. Farrar, Nucl. Phys. Proc. Suppl. {\bf 43}, 312 (1995);
M. E. Shaposhnikov, Phys. Lett. B {\bf 277},324 (1992); 
G. R. Farrar and M. E. Shaposhnikov, Phys. Rev. Lett. {\bf 70}, 2833 (1993). 

\bibitem{trodden} M. Trodden, hep-ph/9803479 (to appear in Rev. Mod. Phys.); 
A. Riotto and M. Trodden, hep-ph/9901362.   

\bibitem{riotto} A. Riotto, hep-ph/9807454. 

\bibitem{joyce} M. Joyce, T. Prokopec and N. Turok, Phys. Lett. B {\bf 338}, 269 (1994); 
{\em ibid.} B {\bf 339}, 312 (1994); Phys. Rev. Lett. {\bf 75}, 1695 (1994); 
Phys. Rev. D {\bf 53}, 2930 (1996); 
{\em ibid.} D {\bf 53}, 2958 (1996). 

\bibitem{moore} G. D. Moore and T. Prokopec, Phys. Rev. D {\bf 52}, 7182 (1995). 

\bibitem{klimov} V. V. Klimov, Sov. J. Nucl. Phys. {\bf 33}, 934 (1981); 
O. K. Kalashnikov and V. V. Klimov, Sov. J. Nucl.
Phys. {\bf 31}, 699 (1980); 
U. Heinz, Ann. Phys. (N. Y.) {\bf 161}, 48 (1985); {\em ibid.} {\bf 168}, 148 (1986).
 

\bibitem{weldon1} H. A. Weldon, Phys. Rev. D {\bf 26}, 1394 (1982); 
{\em ibid.} D {\bf 26}, 2789 (1982); 
{\em ibid.} D {\bf 40}, 2410 (1989); Physica A {\bf 158}, 169 (1989). 

\bibitem{pisa} R. D. Pisarski, Phys. Rev. Lett. {\bf 63}, 1129 (1989);
Nucl. Phys. {\bf A525}, 175 (1991); Physica A {\bf 158}, 146 (1989). 


\bibitem{blaizot1} J.-P. Blaizot, in {\em Proceedings of the Fourth Summer School 
and Symposium on Nuclear Physics}, edited by D. P. Min and M. Rho 
(World Scientific, Singapore, 1991).

\bibitem{kapusta} J. I. Kapusta, {\em Finite Temperature Field Theory} 
(Cambridge University Press, 1989).



\bibitem{lebellac}  M. Le Bellac, {\em Thermal Field Theory}
(Cambridge University Press, 1996).  

\bibitem{vilja1} A. Riotto and I. Vilja, Phys. Lett. B {\bf 402}, 314 (1997). 

\bibitem{holo} T. Holopainen, J. Maalampi, J. Sirkka and I. Vilja, Nucl. Phys. {\bf B473}, 173 (1996). 

\bibitem{davo} H. Davoudiasl and E. Westphal, Phys. Lett. B {\bf 432}, 128 (1998).

\bibitem{htl} R. D. Pisarski, Physica A {\bf 158}, 146 (1989);
Phys. Rev. Lett. {\bf 63}, 1129 (1989); 
E. Braaten and R. D. Pisarski, Nucl. Phys. {\bf B337}, 569 (1990);
{\em ibid.} {\bf B339}, 310 (1990); 
R. D. Pisarski, Nucl. Phys. {\bf B309}, 476 (1988);
J. Frenkel and J. C. Taylor, Nucl. Phys. {\bf B334}, 199 (1990);
J. C. Taylor and S. M. H. Wong, Nucl. Phys. {\bf B346}, 115 (1990).

\bibitem{brapis} E. Braaten and R. D. Pisarski, Phys. Rev. D {\bf 46}, 1829 (1992).

\bibitem{pisarski} R. D. Pisarski, Phys. Rev. D {\bf 47}, 5589 (1993). 

\bibitem{rebhan1} F. Flechsig, A. K. Rebhan and H. Schulz, Phys. Rev. D {\bf 52}, 2994 (1995). 


\bibitem{iancu} J.-P. Blaizot and E. Iancu, Phys. Rev. Lett. {\bf 76}, 3080 (1996); 
Phys. Rev. D {\bf 55}, 973 (1997); {\em ibid.} D {\bf 56}, 7877 (1997).  

\bibitem{boyrgir} D. Boyanovsky, H. J. de Vega, R. Holman, M. Simionato, hep-ph/9809346
(to appear in Phys. Rev. D).

\bibitem{rebhan} U. Kraemmer, A. Rebhan and H. Schulz, Ann. Phys. (N.Y.)
{\bf 238}, 286 (1995). 

\bibitem{fermion}D. Boyanovsky, H. J. de Vega, D.-S. Lee, Y. J. Ng and 
S.-Y. Wang, Phys. Rev. D {\bf 59}, 105001 (1999).

\bibitem{wagner} M. Carena, M. Quiros and C. E. M. Wagner, Phys. Lett. {\bf B380}, 81 (1996).

\bibitem{quiros} M. Quiros, hep-ph/9901312 (unpublished).

\bibitem{meta} D. Metaxas and E. Weinberg, Phys. Rev. {\bf D53}, 836 (1996).

\bibitem{wu} E. Weinberg and A. Wu, Phys. Rev. {\bf D36}, 2472 (1987). 

\bibitem{boyaveff}  D. Boyanovsky, W. Loinaz and  R.S. Willey, Phys.Rev. {\bf D57} 100,  (1998); 
D. Boyanovsky, D. Brahm, R. Holman, D.-S. Lee, Phys.Rev. {\bf D54} (1996) 1763. 

\bibitem{ctp} J. Schwinger, J. Math. Phys. {\bf 2}, 407 (1961); 
K. T. Mahanthappa, Phys. Rev. {\bf 126}, 329 (1962); 
P. M. Bakshi and K. T. Mahanthappa, J. Math. Phys. {\bf 41}, 12 (1963);   
L. V. Keldysh, JETP  {\bf 20}, 1018 (1965); 
K. Chou, Z. Su, B. Hao And L. Yu, Phys. Rep. {\bf 118}, 1 (1985); 
A. Niemi and G. Semenoff, Ann. of Phys. (NY) {\bf 152}, 105 (1984); 
N. P. Landsmann and C. G.  van Weert, Phys. Rep. {\bf 145}, 141 (1987); 
E. Calzetta and B. L. Hu, Phys. Rev. D {\bf 35}, 495 (1987); 
{\em ibid.}  D {\bf 37}, 2878  (1988); 
J. P. Paz, Phys. Rev. D {\bf 41}, 1054 (1990); {\em ibid.} D {\bf 42}, 529 (1990).  

\bibitem{disip}D. Boyanovsky, H. J. de Vega and R. Holman, in {\em Proceedings of
the Second Paris Cosmology Colloquium, Observatoire de Paris, June 1994},
edited by H. J. de Vega and N. Sanchez (World Scientific, Singapore, 1995);
D. Boyanovsky, H. J. de Vega and R. Holman, in {\em
Advances in Astrofundamental Physics, Erice Chalonge Course}, edited by N. Sanchez and
A. Zichichi, (World Scientific, Singapore, 1995); 
D. Boyanovsky, H. J. de Vega, R. Holman, D.-S. Lee and A. Singh,
Phys. Rev. D {\bf 51},  4419 (1995);
D. Boyanovsky, H. J. de Vega, R. Holman and D.-S. Lee, Phys. Rev. D 
{\bf 52}, 6805 (1995).

\bibitem{linon} D. Boyanovsky, M. D'Attanasio, H. J. de Vega and R. Holman, 
Phys. Rev. D {\bf 54}, 1748 (1996); D. Boyanovsky, M. D'Attanasio,
H. J. de Vega, R. Holman and D.-S. Lee, Phys. Rev. D {\bf 52}, 6805 (1995). 


\bibitem{conversation} We thank Robert Pisarski and Edmond Iancu for their
 explanation on this point. 

\bibitem{kobes} R. Baier and R. Kobes, Phys. Rev. D {\bf 50}, 5944 (1994).

\bibitem{pilon} S. Peign\'e, E. Pilon and D. Schiff, Z. Phys. {\bf C60},
455, (1993). 

   










\end{thebibliography}
\end{document}